\documentstyle[11pt,bezier]{article}
\newlength{\awidth}
\newlength{\aheight}
\newlength{\uswidth}
\newlength{\usheight}
\newlength{\spacing}
\newlength{\topoff}
\newlength{\margoff}
\newlength{\margin}
\newlength{\hmargin}
\newcounter{fignr}

\begin{document}
\setlength{\baselineskip}{\spacing}
\def\scri{\unitlength=1.00mm
\thinlines
\begin{picture}(3.5,2.5)(3,3.8)
\put(4.9,5.12){\makebox(0,0)[cc]{$\cal J$}}
\bezier{20}(6.27,5.87)(3.93,4.60)(4.23,5.73)
\end{picture}}

\begin{titlepage}
\noindent {\today \hfill USITP 97-13}
\begin{center}
{\Large BLACK HOLES AND WORMHOLES\\ \vspace{1 ex}
IN 2+1 DIMENSIONS}
\vspace{20mm}

{\large Stefan \AA minneborg}$^*$\footnote{Email address:
stefan@vanosf.physto.se}

\vspace{5mm}

{\large Ingemar Bengtsson}$^{**}$\footnote{Email address:
ingemar@vana.physto.se}

\vspace{5mm}

{\large Dieter Brill}$^{***}$\footnote{Email address:
brill@umdhep.umd.edu}

\vspace{5mm}

{\large S\"{o}ren Holst}$^{**}$\footnote{Email address:
holst@vanosf.physto.se}

\vspace{5mm}

{\large Peter Peld\'{a}n}$^{**}$\footnote{Email address:
peldan@vanosf.physto.se}

\vspace{14mm}

*{\sl \ Norra Reals Gymnasium, S-113 55 Stockholm, Sweden}

** {\sl \ Fysikum, Stockholm University, Box 6730, S-113 85
Stockholm, Sweden}

***{\sl \ Department of Physics, University of Maryland,
College Park, MD 20742, USA}

\vspace{20mm}

{\bf Abstract}
\end{center}
A large variety of spacetimes---including the BTZ black
holes---can be obtained by identifying points in
2+1 dimensional anti-de Sitter space by means of a
discrete group of isometries. We consider all such
spacetimes that can be obtained under a restriction
to time symmetric initial data and one asymptotic
region only. The resulting spacetimes are non-eternal
black holes with collapsing wormhole topologies.
Our approach is geometrical, and we discuss in detail:
The allowed topologies, the shape of the event horizons,
topological censorship and trapped curves.
\end{titlepage}
\noindent
{\large I. Introduction.}

\vspace{3ex}

\noindent Black holes in 2+1 dimensions are remarkable
and surprising not only by their
mere existence, but also because of the wide variety and
topologically distinct types of black-hole
spacetimes that can be constructed. All of these exist as
solutions of the sourcefree Einstein equations only if
there is a negative cosmological constant, and hence they
are asymptotically anti-de Sitter
rather than asymptotically flat. We concentrate attention on
such spacetimes that have just one asymptotic region, but a
complex interior topology.

In 3+1 dimensions the simplest (Schwarzschild-Kruskal)
black hole has two asymptotic regions, and
is eternal in the sense that a horizon for each asymptotic region exists
for all times. Black holes with one asymptotic region are typically the
result of gravitational collapse (of gravitational waves in the source-free
case), and are therefore not eternal.  There are also more bizarre black hole
spacetimes that have only one asymptotic region, with a horizon that starts
at some time somewhere in the interior, but in which there is no
collapse of waves from the exterior. An example is
the ``RP$^3$'' geon \cite{Friedman}, a non-orientable quotient space of the
Schwarzschild spacetime. The details of the dynamics of
non-eternal black holes can be rather
difficult to analyze; much of what we know about their horizons
and how it starts comes from numerical work.

By contrast, in 2+1 dimensions all the details of the black hole solutions
can be analyzed exactly. This is of course due to the absence of local
gravitational degrees of freedom (gravitational waves); but global and
topological degrees of freedom still exist in 2+1 dimensions---in particular
we will see that there is an analog of the black holes with wormhole
topology. Thus, in 2+1
dimensions a great variety of simple models allow detailed investigation,
for example of the collapse and of the horizon.

In this paper we show the construction of 2+1 dimensional black holes with a
single asymptotic
region and arbitrary spatial topology in the interior. By spatial topology
we mean the 2-dimensional topology of spacelike surfaces $\Sigma$ that
foliate the spacetime. (They are not Cauchy surfaces since the spacetime
is modeled on anti-de Sitter space, which is not globally hyperbolic.)
The topology of
such 2-dimensional surfaces is classified by genus and orientability,
and we will see that all topologies with genus $\geq 1$ are allowed.
We will give a complete account of trapped surfaces and the
shape of the event horizon for the simplest wormhole topology, and
indicate what the situation is for the more complicated cases.

In section II we recall the essential features of a single (non-rotating)
2+1 dimensional black hole, the ``BTZ black hole.'' In section
III we introduce a particular point of view \cite{Brill1,Steif}
which enables us to construct a variety of such black holes.
Section IV introduces simple ``building blocks'', out of which
black holes of all interior topologies can be constructed. Section
V discusses the spacetimes that result, with particular attention to
the exterior and to the event horizon. Section VI is devoted
to topological censorship. Section VII
focuses on the trapped surfaces and apparent horizons of these
spacetimes. Two appendices describe the
coordinate systems that we use to visualize the constructions.

\vspace{5ex}
\noindent
{\large II. The Geometry of the Simplest Black Hole Solution.}

\vspace{3ex}

\noindent The 2+1 dimensional black hole found by Ba\~nados,
Teitelboim and Zanelli \cite{BHTZ} is
easily described in terms of Schwarzschild-like coordinates,

\begin{equation} ds_{BTZ}^2 = -\left(-M+(r/\ell)^2\right)dt^2 +
{dr^2\over\left(-M+(r/\ell)^2\right)} + r^2d\phi^2\ . \end{equation}

\noindent (Here we have confined attention to the case without
angular momentum, because spinning black holes are difficult to
handle with the methods that we will use.)
To cover the complete spacetime we need to
allow $r$ to vary between 0 and $\infty$, with the usual Kruskal-type
analytic extension across the lightcone $r=M$. The coordinate $\phi$ is
understood to have periodicity $2\pi$; if it instead is allowed to range
over all of ${\rm I\!R}$, the metric describes (a part of) anti-de Sitter
(adS) space. To understand the resulting
global geometries more directly, other representations are useful.

Since all known 2+1-dimensional black hole geometries are
quotient spaces of adS
space under a group of isometries, we first consider a representation of
adS space itself.
We introduce a flat 4-dimensional space with a metric of somewhat unusual
signature,

\begin{equation} ds^2 = -dT^2 - dU^2 + dX^2 + dY^2 \ . \end{equation}

\noindent Then the 2+1-dimensional adS space can be embedded in this flat space
as the hyperboloidal surface

\begin{equation} -T^2 - U^2 + X^2 + Y^2 = -\ell^2 \ . \end{equation}

\noindent (We will set the scale factor $\ell^2 = 1$ from
now on.) This surface is periodic in time,
so true adS space would be the covering space of this surface. This
distinction is however immaterial for the black holes, which involve less
than one period of the surface. Indeed the covering space of our
black hole spacetimes is a proper subset of the hyperboloid.

One way to describe  2+1-dimensional black holes is to specify the isometry
group by which points in adS space are to be identified.
Alternatively we can choose
a fundamental region of the isometry and describe the spacetime by the
identifications of the region's boundaries implied by the isometry.
The isometries are the rigid motions SO(2,2) that leave the origin
of the embedding  space fixed and its metric invariant.
These isometries map adS space (3) into itself.

The discrete group of transformations corresponding to a BTZ black hole
is generated, for example, by the ``Lorentz-type'' adS transformation

\begin{equation}
\begin{array}{lr}
 T'= T\cosh(2\pi\sqrt{M}) + Y\sinh(2\pi\sqrt{M}) & U'=U  \\
 Y'= T\sinh(2\pi\sqrt{M}) + Y\cosh(2\pi\sqrt{M}) & X'=X  .
\end{array}
\end{equation}

\noindent A possible fundamental region is therefore given by
the (double) wedge

\begin{equation}  -T\tanh(\pi\sqrt{M}) \leq Y \leq T\tanh(\pi\sqrt{M})
\end{equation}

\noindent whose boundaries are to be identified according to the
transformation (4). The edges of the wedge are the fixed point sets
$T=0$, $Y=0$; these are considered as singular points in the BTZ
geometry. The covering space of the BTZ black hole is then given by
a subset of adS space in which the Killing vector $J_{TY} =
T\partial_Y + Y\partial_T$ (which
generates the discrete isometry) is spacelike.

In addition to BTZ coordinates, there are two other convenient sets
of coordinates within adS space called ``stereographic'' and
``sausage'' coordinates.  The stereographic coordinates
express the metric in a form which is manifestly conformally
flat, and are therefore very helpful for visualizing lightlike surfaces.
The sausage coordinates foliate adS with Poincar\'{e} disks
and have the advantages that
they are static and cover the whole manifold. Sausage coordinates are
uniquely defined and make the time-symmetry manifest, whereas
stereographic coordinates can be ``centered'' on various points of
the spacetime (antipodal to the center of projection). Both of these
coordinate systems
have the virtue that they can bring points at infinity to finite
coordinate values---they provide us with
a kind of 3-dimensional Penrose diagrams.

In this paper we will not use intrinsic coordinates to carry
out calculations, but we will frequently draw pictures to visualize
our constructions. In these pictures either stereographic or
sausage coordinates will be used.  For the convenience of the
reader all details about the coordinate systems are given in the
appendix (they can also be found in previous publications
\cite{Brill2} \cite{Holst}). The resulting pictures of the
fundamental region of the BTZ black hole are given in figure 1.

When the fundamental region is given it is easy
to identify the event horizon. It is simply the boundary of that
part of the fundamental region which cannot be seen from its
conformal boundary (\scri). Figure 1 shows this construction for
the BTZ solution, and we see that this spacetime is a
genuine black hole since indeed only a part of the fundamental
region can be reached by past directed null geodesics from \scri.
The fundamental region always has the
property that points that are left fixed by some element of the
group can occur (if at all) only on its boundary, where they
will give rise to ``corners'' or ``folds''.
In figure 1b, which covers the
entire fundamental region, there are two lines
of fixed points, one in the future and one in the past. A fixed
point gives rise to a singularity in the quotient space.
In spacetimes with Euclidean signature only conical singularities
arise in this way, but the singularities of the BTZ solutions
are of a different nature.
Indeed, at the cost of admitting closed timelike curves,
it is possible to extend the solution analytically beyond these
singularities to some extent, to see this,
note that the part of the solution which is covered by the stereographic
coordinates is conformal to 2+1 dimensional Misner space
\cite{Misner}. By means of stereographic coordinates centered on
different points we can cover the entire BTZ black hole with
two Misner spaces. The singularity of a Misner
space is well understood, and the singularities of the BTZ
solution have the same properties. In this paper we simply
refer to them as ``singularities'', and our only concern is to
ensure that they are hidden behind an event horizon (unless
they occur at the beginning of the universe).
\vspace{1.75cm}
\unitlength 0.50mm
\noindent
\parbox{\textwidth}{
\begin{center}
\begin{picture}(218,145.67)
\thicklines
\bezier{10}(60.20,100.00)(60.20,103.53)(66.14,105.48)
\bezier{10}(66.14,105.48)(71.09,107.44)(80.00,107.84)
\bezier{10}(60.20,100.00)(60.20,96.47)(66.14,94.52)
\bezier{10}(66.14,94.52)(71.09,92.16)(80.00,92.16)
\bezier{10}(99.80,100.00)(99.80,103.53)(93.86,105.48)
\bezier{10}(93.86,105.48)(88.91,107.84)(80.00,107.84)
\bezier{10}(99.80,100.00)(99.80,96.47)(93.86,94.52)
\bezier{10}(93.86,94.52)(88.91,92.16)(80.00,92.16)
\bezier{320}(63.67,15.50)(72.67,78.67)(66.00,94.46)
\bezier{312}(66.00,94.46)(57.00,76.00)(18.00,33.00)
\bezier{50}(94.00,105.67)(89.67,88.00)(98.33,64.67)
\bezier{25}(19.00,48.00)(45.00,72.67)(52.33,82.33)
\bezier{15}(52.33,82.33)(60.00,93.00)(60.00,100.00)
\thinlines
\bezier{268}(133.00,46.70)(97.67,82.00)(80.00,81.00)
\bezier{268}(27.00,47.00)(62.33,82.00)(80.00,81.00)
\bezier{244}(71.33,18.67)(81.33,68.33)(89.67,71.00)
\bezier{70}(89.67,71.00)(99.33,76.00)(134.66,43.80)
\thicklines
\bezier{160}(142.00,47.00)(115.00,72.67)(107.67,82.33)
\bezier{15}(107.67,82.33)(100.00,93.00)(100.00,100.00)
\thinlines
\bezier{275}(25.67,36.00)(58.00,75.00)(69.00,79.00)
\bezier{40}(69.00,79.00)(80.33,83.00)(91.00,61.67)
\put(94.00,105.89){\line(-1,-1){29.00}}
\put(94.00,105.00){\line(3,-5){45.00}}
\bezier{68}(76.50,88.33)(70.83,83.17)(80.00,80.00)
\bezier{120}(80.00,80.00)(94.17,76.00)(108.50,80.67)
\bezier{92}(102.33,50.00)(117.00,51.00)(124.00,55.33)
\bezier{100}(115.00,23.22)(128.33,25.22)(139.22,29.89)
\thicklines
\bezier{112}(94,105.3)(96.11,94.78)(107.67,82.33)
\put(63.44,15.22){\line(5,2){46.67}}
\thinlines
\put(110.11,33.89){\line(5,2){21.56}}
\thicklines
\put(131.67,42.51){\line(5,2){10.33}}
\thinlines
\bezier{80}(75.22,80.56)(83.89,80.78)(89.89,71.22)
\bezier{216}(89.89,71.22)(97.89,59.44)(115.00,23.67)
\bezier{28}(85.66,97.81)(83.88,95.47)(87.55,93.92)
\bezier{56}(87.55,93.92)(93.99,91.59)(100.55,94.14)
\bezier{88}(93.97,64.84)(106.37,65.37)(115.57,68.97)
\bezier{92}(108.93,37.04)(121.31,38.47)(131.07,42.75)
\thicklines
\put(18.10,33.10){\line(5,2){50.00}}
\bezier{30}(68.33,53.10)(83.57,58.81)(98.50,64.29)
\bezier{20}(19.09,47.88)(12.42,40.00)(19.09,31.82)
\bezier{32}(19.09,31.82)(30.30,20.00)(63.64,15.15)
\bezier{55}(63.64,15.15)(119.70,12.12)(139.39,29.70)
\bezier{23}(139.39,29.70)(146.97,36.97)(142.00,47.00)
\bezier{25}(19.00,47.78)(28.72,58.33)(54.28,63.06)
\bezier{30}(54.28,63.06)(77.33,66.39)(99.56,64.17)
\bezier{27}(99.56,64.17)(121.22,61.39)(137.33,51.67)
\thinlines
\bezier{35}(27.00,47.00)(24.50,43.82)(24.64,40.51)
\bezier{5}(24.64,40.51)(24.60,38.26)(25.92,36.15)
\bezier{188}(25.92,36.15)(33.56,23.50)(64.13,18.72)
\bezier{6}(64.13,18.72)(67.72,18.40)(71.31,18.21)
\bezier{180}(71.31,18.21)(94.64,17.18)(114.90,23.33)
\bezier{25}(114.90,23.33)(132.85,29.23)(134.64,37.18)
\bezier{44}(134.64,37.18)(136.00,41.79)(133.10,46.41)
\bezier{40}(133.10,46.41)(122.85,59.23)(85.92,61.79)
\bezier{45}(85.92,61.79)(39.49,61.03)(27.00,47.00)
\thicklines
\put(66.00,94.56){\line(5,2){28.11}}
\thinlines
\put(108.33,110.67){\vector(-3,-1){12.67}}
\put(109.00,110.67){\makebox(0,0)[lc]{P}}
\put(80.50,99.5){\makebox(0,0)[cc]{*}}
\put(87.50,102.3){\makebox(0,0)[cc]{*}}
\put(72.50,96.4){\makebox(0,0)[cc]{*}}
\put(45.00,32.00){\vector(1,0){0.2}}
\bezier{132}(23.00,18.00)(28.00,33.00)(45.00,32.00)
\put(23.00,15.00){\makebox(0,0)[cc]{\small Initial Data Surface}}
\put(23.00,9.00){\makebox(0,0)[ct]{\small $(U=0)$}}
\put(115.00,94.00){\line(-3,-1){11.50}}
\put(97.73,88.40){\vector(-3,-1){0.2}}
\bezier{6}(102.80,90.00)(100.27,89.20)(97.73,88.40)
\put(115.50,94.00){\makebox(0,0)[lc]{\scri}}
\bezier{80}(210.00,25.00)(210.00,19.17)(200.00,16.25)
\bezier{80}(200.00,16.25)(190.00,13.34)(180.00,16.25)
\bezier{80}(180.00,16.25)(170.00,19.17)(170.00,25.00)
\bezier{80}(210.00,75.00)(210.00,70.20)(200.00,67.80)
\bezier{80}(200.00,67.80)(190.00,65.41)(180.00,67.80)
\bezier{80}(180.00,67.80)(170.00,70.20)(170.00,75.00)
\bezier{32}(196.00,67.11)(194.78,68.56)(198.00,70.01)
\bezier{44}(198.00,70.01)(202.11,71.26)(206.00,70.01)
\bezier{32}(184.00,82.89)(185.22,81.44)(182.00,79.99)
\bezier{44}(182.00,79.99)(177.89,78.74)(174.00,79.99)
\bezier{80}(210.00,125.00)(210.00,129.15)(200.00,131.22)
\bezier{80}(200.00,131.22)(190.00,133.29)(180.00,131.22)
\bezier{80}(180.00,131.22)(170.00,129.15)(170.00,125.00)
\bezier{80}(210.00,125.00)(210.00,120.85)(200.00,118.78)
\bezier{80}(200.00,118.78)(190.00,116.71)(180.00,118.78)
\bezier{80}(180.00,118.78)(170.00,120.85)(170.00,125.00)
\put(170.00,25.00){\line(0,1){100.00}}
\put(210.00,25.00){\line(0,1){100.00}}
\put(80.00,0.00){\makebox(0,0)[cc]{a}}
\put(190.00,0.00){\makebox(0,0)[cc]{b}}
\thicklines
\put(174.50,120.50){\line(3,1){29.50}}
\put(174.00,19.00){\line(5,2){31.00}}
\put(190.00,24.60){\makebox(0,0)[cc]{*}}
\put(200.00,28.60){\makebox(0,0)[cc]{*}}
\put(180.00,20.46){\makebox(0,0)[cc]{*}}
\put(190.00,124.7){\makebox(0,0)[cc]{*}}
\put(180.00,121.6){\makebox(0,0)[cc]{*}}
\put(200.00,128.10){\makebox(0,0)[cc]{*}}
\bezier{536}(174.50,120.50)(218.00,64.33)(174.00,19.00)
\bezier{148}(204.00,130.00)(211.00,115.33)(210.00,95.00)
\bezier{128}(208.00,56.00)(204.00,70.33)(208.00,85.00)
\bezier{80}(208.00,85.00)(210.00,95.00)(210.00,95.00)
\bezier{88}(208.00,56.00)(210.00,48.00)(210.00,45.00)
\bezier{44}(210.00,45.00)(210.00,37.00)(205.00,31.50)
\bezier{34}(204.00,130.00)(192.67,116.33)(187.00,101.00)
\bezier{228}(187.00,101.00)(178.00,76.33)(193.00,50.00)
\bezier{23}(193.00,50.00)(197.00,41.33)(205.00,31.50)
\bezier{148}(174.00,19.00)(169.00,34.67)(170.00,55.00)
\bezier{128}(172.00,96.00)(176.00,79.67)(172.00,65.00)
\bezier{80}(172.00,65.00)(170.00,55.00)(170.00,55.00)
\bezier{88}(172.00,96.00)(170.00,102.00)(170.00,107.00)
\bezier{44}(170.00,108.00)(170.00,113.00)(174.50,120.50)
\thinlines
\put(162.00,116.00){\vector(3,1){11.00}}
\put(161.00,117.00){\makebox(0,0)[cc]{P}}
\put(211.00,75.00){\makebox(0,0)[lc]{$t=0$}}
\bezier{40}(210.00,25.00)(210.00,28.50)(205.00,31.50)
\bezier{19}(205.00,31.50)(195.00,36.00)(187.00,35.00)
\bezier{17}(187.00,35.00)(176.50,34.00)(171.50,29.00)
\bezier{20}(171.50,29.00)(170.00,27.30)(170.00,25.00)
\bezier{32}(170.00,75.00)(169.60,77.80)(174.00,80.07)
\bezier{10}(174.00,80.07)(178.53,82.33)(183.60,83.00)
\bezier{40}(183.60,83.00)(188.67,83.67)(193.87,83.27)
\bezier{14}(193.87,83.27)(201.60,82.47)(206.80,79.53)
\bezier{24}(206.80,79.53)(209.47,78.33)(210.00,75.00)
\put(51.00,89.50){\vector(3,1){14.00}}
\put(50.00,89.50){\makebox(0,0)[rc]{P}}
\put(218.00,135.00){\vector(-3,-1){12.00}}
\put(219.00,135.00){\makebox(0,0)[lc]{P}}
\end{picture}\\
\end{center}
\bigskip

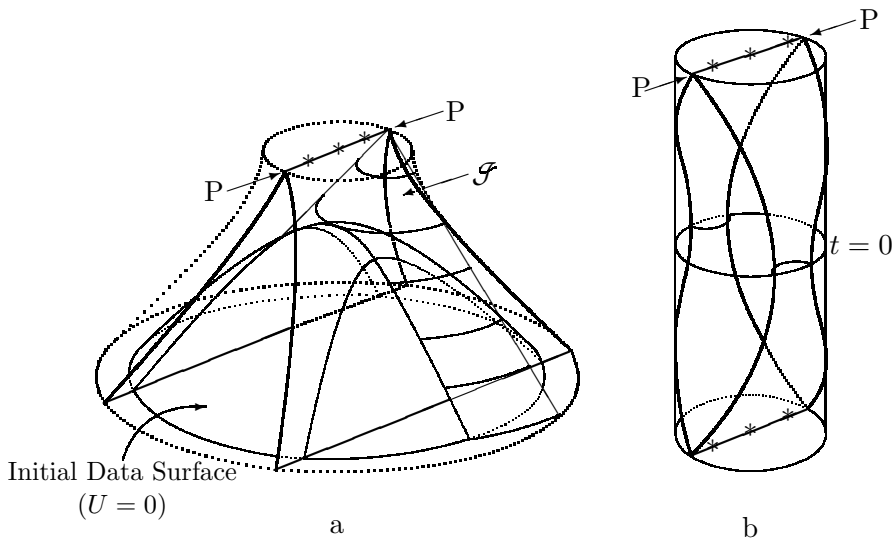
\noindent {\small Fig.~1: The fundamental region of
the BTZ black hole within adS space, indicated by the heavy outline.
The left and right boundaries of this region are to be identified. The
identification generates two lines of singularities as indicated by
``barbed wire'' lines, containing crosses. The singularity meets infinity at
points such as those denoted by P.\\
(a) Stereographic coordinates centered on
the future singularity cover only somewhat more than half of the
fundamental region. The complementary patch, covering the past
singularity, would be a mirror image of this figure about a horizontal plane.
Infinity is represented by the dotted hyperboloid.
The event horizon of one of the two parts of null infinity is given by the
backwards light cone (horizontal stripes) from the end point P of that
null infinity (\scri).\\
(b) In sausage coordinates infinity is represented by a cylinder.
These coordinates cover all of the fundamental region, and the
symmetry about the initial data surface $t=0$ is manifest.
The same surface occurs as the $U = 0$ hyperboloid in the figure
on the left.}
}

\newpage
\noindent
{\large III. Constructing Black Holes from Initial Data.}

\vspace{3ex}

\noindent How can we generalize the BTZ solution?
Let us for the moment set aside the spacetime properties
of our black holes, and instead adopt an initial value point of view
\cite{Brill1}\cite{Steif}. Since adS space is not globally hyperbolic
the future of a set of initial values is not determined
by Cauchy data alone, but this does not prevent us from singling
out a moment in time and studying the spatial geometry at that
moment. Specifically, let us choose the spatial surface defined
by

\begin{equation} U = 0 \hspace{1cm} - T^2 + X^2 + Y^2 = - 1
\hspace{1cm} T > 0
\ . \end{equation}

\noindent This is one sheet of a two dimensional hyperboloid
embedded in the flat 3-dimensional Minkowski space that is defined by $U = 0$.
(As such it would seem to have non-vanishing extrinsic curvature; but
note that as a subspace of the 3-dimensional surface (3) its
extrinsic curvature vanishes!). It
is well known\footnote{For details see, for example, reference
\cite{Mathematics}.} that the intrinsic geometry on such a surface
is a model for Lobachevsky's hyperbolic geometry H$^2$, and that it
is equivalent (under stereographic projection) to that of
the Poincar\'{e} disk. On the latter, spatial geodesics are
represented by circular arcs that are orthogonal
to the boundary of the unit disk; angles are
represented accurately in such a picture, while distances are
distorted. Because the extrinsic curvature of the surface (6) vanishes,
the isometry which gave rise to the BTZ black hole is also
an isometry of our spatial slice; indeed it acts like a hyperbolic
M\"{o}bius transformation---which we will call a ``transvection''.

\vspace{1cm}
\noindent
\parbox{\textwidth}{
\begin{center}
\unitlength 0.85mm
\linethickness{0.4pt}
\begin{picture}(150.00,30.00)(5,0)
\bezier{10}(10.00,20.00)(10.00,24.14)(12.93,27.07)
\bezier{10}(12.93,27.07)(15.86,30.00)(20.00,30.00)
\bezier{10}(30.00,20.00)(30.00,24.14)(27.07,27.07)
\bezier{10}(27.07,27.07)(24.14,30.00)(20.00,30.00)
\bezier{10}(10.00,20.00)(10.00,15.86)(12.93,12.93)
\bezier{10}(12.93,12.93)(15.86,10.00)(20.00,10.00)
\bezier{10}(30.00,20.00)(30.00,15.86)(27.07,12.93)
\bezier{10}(27.07,12.93)(24.14,10.00)(20.00,10.00)
\bezier{64}(13.62,12.19)(11.14,19.90)(13.62,27.81)
\bezier{44}(13.62,12.19)(18.19,14.57)(18.57,20.00)
\bezier{32}(13.62,12.19)(16.76,10.67)(20.67,12.38)
\bezier{40}(20.67,12.38)(25.14,14.86)(25.43,20.00)
\bezier{44}(13.62,27.81)(18.19,25.43)(18.57,20.00)
\bezier{32}(13.62,27.81)(16.76,29.33)(20.67,27.62)
\bezier{40}(20.67,27.62)(25.14,25.14)(25.43,20.00)
\put(12.38,20.00){\vector(0,1){0.19}}
\put(18.57,20.00){\vector(0,1){0.19}}
\put(25.43,20.00){\vector(0,1){0.19}}
\put(20.00,4.00){\makebox(0,0)[ct]{a}}
\bezier{10}(40.00,20.00)(40.00,24.14)(42.93,27.07)
\bezier{10}(42.93,27.07)(45.86,30.00)(50.00,30.00)
\bezier{10}(60.00,20.00)(60.00,24.14)(57.07,27.07)
\bezier{10}(57.07,27.07)(54.14,30.00)(50.00,30.00)
\bezier{10}(40.00,20.00)(40.00,15.86)(42.93,12.93)
\bezier{10}(42.93,12.93)(45.86,10.00)(50.00,10.00)
\bezier{10}(60.00,20.00)(60.00,15.86)(57.07,12.93)
\bezier{10}(57.07,12.93)(54.14,10.00)(50.00,10.00)
\put(44.22,20.00){\circle{8.00}}
\bezier{32}(39.89,20.00)(39.89,22.94)(41.98,25.02)
\bezier{32}(41.98,25.02)(44.06,27.11)(47.00,27.11)
\bezier{32}(54.11,20.00)(54.11,22.94)(52.02,25.02)
\bezier{32}(52.02,25.02)(49.94,27.11)(47.00,27.11)
\bezier{32}(39.89,20.00)(39.89,17.06)(41.98,14.98)
\bezier{32}(41.98,14.98)(44.06,12.89)(47.00,12.89)
\bezier{32}(54.11,20.00)(54.11,17.06)(52.02,14.98)
\bezier{32}(52.02,14.98)(49.94,12.89)(47.00,12.89)
\put(48.38,20.00){\vector(0,1){0.19}}
\put(54.13,20.00){\vector(0,1){0.19}}
\put(50.00,5.11){\makebox(0,0)[ct]{b}}
\bezier{10}(70.00,20.00)(70.00,24.14)(72.93,27.07)
\bezier{10}(72.93,27.07)(75.86,30.00)(80.00,30.00)
\bezier{10}(90.00,20.00)(90.00,24.14)(87.07,27.07)
\bezier{10}(87.07,27.07)(84.14,30.00)(80.00,30.00)
\bezier{10}(70.00,20.00)(70.00,15.86)(72.93,12.93)
\bezier{10}(72.93,12.93)(75.86,10.00)(80.00,10.00)
\bezier{10}(90.00,20.00)(90.00,15.86)(87.07,12.93)
\bezier{10}(87.07,12.93)(84.14,10.00)(80.00,10.00)
\put(80.03,20.00){\circle{6.67}}
\bezier{35}(73.24,19.97)(73.24,22.77)(75.23,24.75)
\bezier{35}(75.23,24.75)(77.21,26.74)(80.01,26.74)
\bezier{35}(86.78,19.97)(86.78,22.77)(84.80,24.75)
\bezier{35}(84.80,24.75)(82.81,26.74)(80.01,26.74)
\bezier{35}(73.24,19.97)(73.24,17.17)(75.23,15.18)
\bezier{35}(75.23,15.18)(77.21,13.20)(80.01,13.20)
\bezier{35}(86.78,19.97)(86.78,17.17)(84.80,15.18)
\bezier{35}(84.80,15.18)(82.81,13.20)(80.01,13.20)
\put(83.50,20.00){\vector(0,1){0.19}}
\put(86.85,20.00){\vector(0,1){0.19}}
\put(80.03,4.00){\makebox(0,0)[ct]{c}}
\bezier{10}(100.00,20.00)(100.00,24.14)(102.93,27.07)
\bezier{10}(102.93,27.07)(105.86,30.00)(110.00,30.00)
\bezier{10}(120.00,20.00)(120.00,24.14)(117.07,27.07)
\bezier{10}(117.07,27.07)(114.14,30.00)(110.00,30.00)
\bezier{10}(100.00,20.00)(100.00,15.86)(102.93,12.93)
\bezier{10}(102.93,12.93)(105.86,10.00)(110.00,10.00)
\bezier{10}(120.00,20.00)(120.00,15.86)(117.07,12.93)
\bezier{10}(117.07,12.93)(114.14,10.00)(110.00,10.00)
\bezier{32}(103.62,12.19)(101.14,19.90)(103.62,27.81)
\bezier{22}(103.62,12.19)(108.19,14.57)(108.57,20.00)
\bezier{16}(103.62,12.19)(106.76,10.67)(110.67,12.38)
\bezier{20}(110.67,12.38)(115.14,14.86)(115.43,20.00)
\bezier{22}(103.62,27.81)(108.19,25.43)(108.57,20.00)
\bezier{16}(103.62,27.81)(106.76,29.33)(110.67,27.62)
\bezier{20}(110.67,27.62)(115.14,25.14)(115.43,20.00)
\put(102.38,20.00){\vector(0,1){0.19}}
\put(108.57,20.00){\vector(0,1){0.19}}
\put(115.43,20.00){\vector(0,1){0.19}}
\thicklines
\bezier{44}(100.14,22.43)(105.86,21.19)(110.14,23.76)
\bezier{28}(110.14,23.76)(113.29,25.29)(115.00,28.62)
\bezier{44}(100.14,17.67)(105.86,18.90)(110.14,16.33)
\bezier{28}(110.14,16.33)(113.29,14.81)(115.00,11.48)
\thinlines
\put(109.00,5.00){\makebox(0,0)[ct]{d}}
\bezier{10}(130.00,20.00)(130.00,24.14)(132.93,27.07)
\bezier{10}(132.93,27.07)(135.86,30.00)(140.00,30.00)
\bezier{10}(150.00,20.00)(150.00,24.14)(147.07,27.07)
\bezier{10}(147.07,27.07)(144.14,30.00)(140.00,30.00)
\bezier{10}(130.00,20.00)(130.00,15.86)(132.93,12.93)
\bezier{10}(132.93,12.93)(135.86,10.00)(140.00,10.00)
\bezier{10}(150.00,20.00)(150.00,15.86)(147.07,12.93)
\bezier{10}(147.07,12.93)(144.14,10.00)(140.00,10.00)
\thicklines
\bezier{80}(131.33,25.00)(140.00,20.33)(148.67,25.00)
\bezier{80}(131.33,15.00)(140.00,19.67)(148.67,15.00)
\thinlines
\bezier{7}(140.00,22.56)(140.00,20.00)(140.00,17.44)
\put(140.00,4.00){\makebox(0,0)[ct]{e}}
\end{picture}
\end{center}

\smallskip
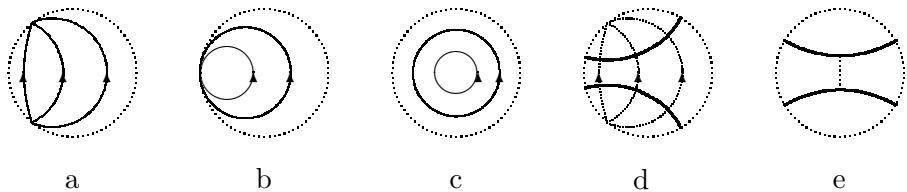
\noindent {\small Fig.~2: Isometries (a-c) and fundamental regions (d,e)
of the Poincar\'e disk.
Continuous isometries are shown by their flow lines.
Discrete isometries are defined by their
fundamental regions in heavy
outline. Infinity is the dotted boundary of the disk. The isometries are
(a) a transvection (with two fixed points on the
boundary of the disk),
(b) a parabolic M\"{o}bius transformation (with
one fixed point on the boundary) and
(c) a rotation (with one fixed
point in the interior, in this case at the center of the disk).\\
(d) A fundamental region of a BTZ black hole at the moment of
time symmetry, obtained by identification under a discrete version of the
isometry shown in (a). \\
(e) The same fundamental region as in (d), after applying a transvection
to move the event horizon (dotted line) to the center of the disk; like the
boundaries of the fundamental region, the event horizon at the moment of time
symmetry is a geodesic.}
}

\vspace{5ex}
Figure 2 is intended to remind the reader about the flow of the
various kinds of isometries that act on the Poincar\'{e} disk.
An interesting fact is that there is one and only
one flow line of a transvection which is also a geodesic.
The figure also shows the intersection of the fundamental region of the
BTZ black hole with the disk; its boundaries are two geodesics.
It is easy to locate the intersection of the event horizon with the
disk, since it is in fact the intersection of two totally geodesic
surfaces, and hence a geodesic on the disk. Since it is also closed,
it must coincide with the unique geodesic flow line of the identifying
transvection.

This construction suggests some obvious generalizations. For example, we can
use a parabolic M\"{o}bius transformation to carry out the
identification. This gives us a ``$M=0$'' (extremal) BTZ black hole (Fig.~3a).
Or we can use a group generated by more than one transvection. As is
well known \cite{Mathematics}, such groups can be defined
by assigning a polygon with $4g$ geodesic sides as a fundamental
region and arranging the identifications so that all corners are
identified as one point.
As long as we arrange for the sum of the polygon's angles to
equal $2{\pi}$ the result is a smooth compact Riemann surface of
genus $g$ (fig.~3b).
In this way we obtain Riemann surfaces of constant
negative curvature for any genus $g \geq 2$; the case of genus one is
excluded since the sum of the angles of a hyperbolic square is necessarily
less than $2{\pi}$, which means that the
resulting torus will have a conical singularity. In the
literature on 2+1 gravity \cite{Deser} conical
singularities are often taken as models of point particles, but
in this paper we adopt the convention that no singularities are
admitted, unless they are hidden by event horizons. Having decided
this we have to face the fact that the torus
does not admit a metric of constant negative curvature. We
can avoid this problem by poking a hole in the torus, and
throwing the boundary of the hole to infinity. The
fundamental region of Figure 3c (with geodesic sides
identified crosswise) indeed gives rise to a smooth
non-compact quotient space of this kind.

\vspace{5mm}
\noindent
\parbox{\textwidth}{
\begin{center}
\unitlength 1.10mm
\linethickness{0.4pt}
\begin{picture}(110.00,30.00)(5,5)
\bezier{11}(10.00,20.00)(10.00,23.75)(12.92,27.08)
\bezier{11}(12.92,27.08)(16.25,30.00)(20.00,30.00)
\bezier{11}(30.00,20.00)(30.00,23.75)(27.08,27.08)
\bezier{11}(27.08,27.08)(23.75,30.00)(20.00,30.00)
\bezier{11}(10.00,20.00)(10.00,16.25)(12.92,12.92)
\bezier{11}(12.92,12.92)(16.25,10.00)(20.00,10.00)
\bezier{11}(30.00,20.00)(30.00,16.25)(27.08,12.92)
\bezier{11}(27.08,12.92)(23.75,10.00)(20.00,10.00)
\bezier{11}(50.00,20.00)(50.00,23.75)(52.92,27.08)
\bezier{11}(90.00,20.00)(90.00,23.75)(92.92,27.08)
\bezier{11}(52.92,27.08)(56.25,30.00)(60.00,30.00)
\bezier{11}(92.92,27.08)(96.25,30.00)(100.00,30.00)
\bezier{11}(70.00,20.00)(70.00,23.75)(67.08,27.08)
\bezier{11}(110.00,20.00)(110.00,23.75)(107.08,27.08)
\bezier{11}(67.08,27.08)(63.75,30.00)(60.00,30.00)
\bezier{11}(107.08,27.08)(103.75,30.00)(100.00,30.00)
\bezier{11}(50.00,20.00)(50.00,16.25)(52.92,12.92)
\bezier{11}(90.00,20.00)(90.00,16.25)(92.92,12.92)
\bezier{11}(52.92,12.92)(56.25,10.00)(60.00,10.00)
\bezier{11}(92.92,12.92)(96.25,10.00)(100.00,10.00)
\bezier{11}(70.00,20.00)(70.00,16.25)(67.08,12.92)
\bezier{11}(110.00,20.00)(110.00,16.25)(107.08,12.92)
\bezier{11}(67.08,12.92)(63.75,10.00)(60.00,10.00)
\bezier{11}(107.08,12.92)(103.75,10.00)(100.00,10.00)
\thicklines
\bezier{70}(10.00,20.00)(14.70,20.00)(19.00,23.00)
\bezier{70}(19.00,23.00)(22.50,26.00)(23.75,29.20)
\bezier{70}(10.00,20.00)(14.70,20.00)(19.00,17.00)
\bezier{70}(19.00,17.00)(22.50,14.00)(23.75,10.80)
\bezier{60}(56.71,28.00)(59.95,24.95)(63.29,28.00)
\bezier{60}(63.29,28.00)(63.57,23.24)(68.05,23.33)
\bezier{60}(51.95,16.76)(55.00,20.00)(51.95,23.33)
\bezier{60}(51.95,23.33)(56.71,23.62)(56.62,28.10)
\bezier{60}(63.29,12.00)(60.05,15.05)(56.71,12.00)
\bezier{60}(56.71,12.00)(56.43,16.76)(51.95,16.67)
\bezier{60}(68.05,23.24)(65.00,20.00)(68.05,16.67)
\bezier{60}(68.05,16.67)(63.29,16.38)(63.38,11.90)
\bezier{40}(94.96,28.50)(96.86,25.81)(100.00,25.81)
\bezier{40}(105.04,28.50)(103.14,25.81)(100.00,25.81)
\bezier{40}(108.50,25.04)(105.81,23.14)(105.81,20.00)
\bezier{40}(108.50,14.96)(105.81,16.86)(105.81,20.00)
\bezier{40}(105.04,11.50)(103.14,14.19)(100.00,14.19)
\bezier{40}(94.96,11.50)(96.86,14.19)(100.00,14.19)
\bezier{40}(91.50,14.96)(94.19,16.86)(94.19,20.00)
\bezier{40}(91.50,25.04)(94.19,23.14)(94.19,20.00)
\thinlines
\bezier{35}(20.50,15.83)(22.10,17.50)(22.10,20.00)
\bezier{35}(22.10,20.00)(22.10,22.50)(20.50,24.00)
\put(59.95,13.52){\vector(0,1){12.86}}
\put(55.57,15.62){\vector(1,1){8.86}}
\put(53.48,20.00){\vector(1,0){12.95}}
\put(20.50,24.00){\vector(-1,1){0.17}}
\put(94.29,20.00){\vector(1,0){11.24}}
\put(100.00,14.29){\vector(0,1){11.24}}
\put(103.14,20.19){\makebox(0,0)[cb]{$a$}}
\put(99.81,23.52){\makebox(0,0)[rc]{$b$}}
\put(91.00,25.00){\makebox(0,0)[rb]{A}}
\put(91.00,14.00){\makebox(0,0)[rc]{C}}
\put(109.00,25.00){\makebox(0,0)[lb]{A}}
\put(109.00,14.00){\makebox(0,0)[lc]{C}}
\put(94.00,29.00){\makebox(0,0)[cb]{B}}
\put(106.00,29.00){\makebox(0,0)[cb]{D}}
\put(94.00,11.00){\makebox(0,0)[ct]{B}}
\put(106.00,11.00){\makebox(0,0)[ct]{D}}
\put(20.00,4.00){\makebox(0,0)[cc]{a}}
\put(60.00,4.50){\makebox(0,0)[cc]{b}}
\put(100.00,4.00){\makebox(0,0)[cc]{c}}
\put(64.45,15.55){\vector(-1,1){8.9}}
\end{picture}
\end{center}

\smallskip
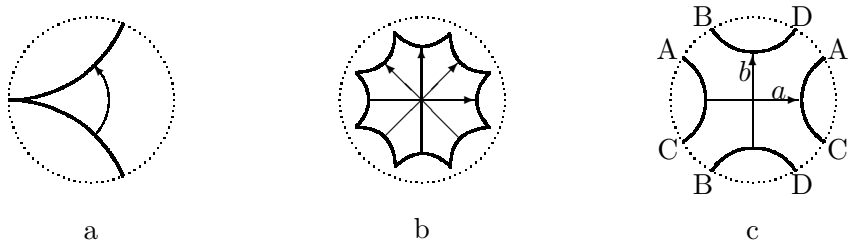
\noindent {\small Fig.~3: Initial data represented by fundamental
regions and identifications (arrows), corresponding to (a) an
extremal black hole, (b)
a compact universe of genus 2, and (c) a wormhole with one asymptotic region.
The latter has only one asymptotic region since the
path $A \rightarrow B \rightarrow C \rightarrow D \rightarrow
A$ is closed.}
}
\vspace{5mm}

To evolve these data in time, we note that the extrinsic
curvature tensor vanishes, so we have a time symmetric initial
value problem; but since adS space is not globally hyperbolic, further
information is needed to specify the spacetime. We bypass this problem
by adopting the rule that the spacetime is given by extending the
action of the discrete isometry group to anti-de Sitter space, and
then we construct the quotient space. This is of course equivalent to
extending analytically beyond the initial data's domain of
dependence. When this procedure is applied
to the initial data in Figure 2d we recover the BTZ black hole.
For the initial data in Figure 3 we obtain respectively
an extremal black hole \cite{BHTZ}, a compact universe which begins
and ends with a singularity \cite{Mess} \cite{Brill2}, and
a new kind of 2+1 black hole that is the subject of this paper.

It may be useful to restate our strategy in spacetime terms
as follows: We are interested in spacetimes that can be obtained
from anti-de Sitter space as the quotient space adS/${\Gamma}$, where
${\Gamma}$ is a discrete group which acts properly discontinuously
on some subset of adS. Such spacetimes were given a thorough
treatment by Mess \cite{Mess}, who classified all the quotient
spacetimes that can be obtained in this way and which have the
topology ${\bf S}\otimes {\rm I\!R}$, where ${\bf S}$ is a closed
spatial surface. For our purposes this is not enough. We want
non-compact spaces (so that infinity can be defined), and we
admit groups with fixed points, but only if
the corresponding singularities in the quotient space are hidden
behind event horizons. To identify such black hole solutions
we need to understand the causal structure of the entire quotient
spacetime. For this reason we restrict our problem by the demand
that ${\Gamma}$ shall transform the surface $U = 0$ into itself,
which means that ${\Gamma}$ belongs to a diagonal $SO(2,1)$
subgroup of $SO(2,2)$.  As we will see this is enough to make
our problem manageable. (The restriction is clearly severe,
in particular it excludes the spinning black hole found by
Ba\~nados, Henneaux, Teitelboim and Zanelli \cite{BHTZ}.
We will comment briefly on this point in the concluding
section.) Even so discrete
groups of this kind are notoriously difficult to describe by
their Lorentz matrix representation. The
most practical way to define a particular discrete group
is to specify a fundamental
region on the  initial Poincar\'{e} disk, $U = 0$.
The properties of adS/${\Gamma}$ can then be found through
an analysis of the fundamental region in adS.
For example, in Fig.~3c
${\Gamma}$ is the discrete group that is generated by the two
transformations $a$ and $b$ that identify opposite sides.
A similar prescription may be
adopted for Fig.~3b, but in the latter
case the generators are subject to a restriction which
ensures that the sum of the angles of the octagon is $2{\pi}$.

Let us give a brief sketch of our new black hole here (to be
followed by a careful analysis in sections V-VII). The
fundamental region in Fig.~3c serves to define the two group
elements $a, b$ that generate ${\Gamma}$. It is then straightforward
to find the fundamental region in anti-de Sitter space; we draw
its spacetime portrait in sausage coordinates in Fig.~4. It
looks like a tent with four openings and totally geodesic sides that
meet at four ``folds'' in the roof of the tent.
The folds are lines of fixed points for certain elements of $\Gamma$.
To see this it must be remembered that ${\Gamma}$ has an infinite number
of elements even though it has only two generators.
The general element transforms a fold into another fold that may lie in
another image of the fundamental region, outside the tent; but there are
particular elements, of the type $aba^{-1}b^{-1}$ that leave a particular
fold fixed. These elements give rise to singularities in the quotient space.
Drawing the picture is a quick way to locate these lines.
It should be kept in mind that in the quotient space, that is after
the identifications according to $a$ and $b$ have been
made, there is really only one asymptotic region, and
similarly there is only one fold in the roof.

In analogy to the case of the BTZ black hole, it is evident
that all of the interior of the fundamental region cannot be seen
using null lines from the openings of the ``tent''.
By definition, the hidden
region is the black hole. Since the folds in the roof cannot be
seen from the openings the resulting singularities are hidden
inside the black hole---no naked singularities will be created.

\vspace{1cm}
\noindent
\parbox{\textwidth}{
\unitlength 0.70mm
\linethickness{0.4pt}
\begin{picture}(97.00,133.29)(-35,5)
\bezier{80}(80.00,25.00)(80.00,30.83)(70.00,33.75)
\bezier{28}(50.00,33.75)(53.50,34.40)(56.50,34.50)
\bezier{28}(70.00,33.75)(66.50,34.40)(63.50,34.50)
\bezier{80}(50.00,33.75)(40.00,30.83)(40.00,25.00)
\bezier{80}(80.00,25.00)(80.00,19.17)(70.00,16.25)
\bezier{80}(70.00,16.25)(60.00,13.34)(50.00,16.25)
\bezier{80}(50.00,16.25)(40.00,19.17)(40.00,25.00)
\bezier{20}(80.00,75.00)(80.00,79.80)(70.00,82.20)
\bezier{20}(70.00,82.20)(60.00,84.59)(50.00,82.20)
\bezier{20}(50.00,82.20)(40.00,79.80)(40.00,75.00)
\bezier{80}(80.00,75.00)(80.00,70.20)(70.00,67.80)
\bezier{80}(70.00,67.80)(60.00,65.41)(50.00,67.80)
\bezier{80}(50.00,67.80)(40.00,70.20)(40.00,75.00)
\bezier{32}(41.00,72.51)(44.50,73.01)(48.00,71.68)
\bezier{44}(48.00,71.68)(51.00,69.97)(48.00,68.35)
\bezier{32}(66.00,67.11)(66.00,69.50)(68.00,70.01)
\bezier{44}(68.00,70.01)(72.11,71.26)(76.00,70.01)
\bezier{36}(53.93,83.00)(54.00,80.60)(51.00,79.63)
\bezier{7}(51.00,79.63)(47.27,79.00)(43.90,80.07)
\bezier{12}(79.00,77.50)(77.93,77.30)(76.60,77.35)
\bezier{15}(76.60,77.35)(67.13,78.73)(72.20,81.67)
\bezier{80}(80.00,125.00)(80.00,129.15)(70.00,131.22)
\bezier{80}(70.00,131.22)(60.00,133.29)(50.00,131.22)
\bezier{80}(50.00,131.22)(40.00,129.15)(40.00,125.00)
\bezier{80}(80.00,125.00)(80.00,120.85)(70.00,118.78)
\bezier{80}(70.00,118.78)(60.00,116.71)(50.00,118.78)
\bezier{80}(50.00,118.78)(40.00,120.85)(40.00,125.00)
\put(40.00,25.00){\line(0,1){100.00}}
\put(80.00,25.00){\line(0,1){100.00}}
\thicklines
\bezier{252}(57.00,41.00)(39.00,71.33)(57.00,92.00)
\bezier{248}(57.00,92.00)(75.00,67.33)(57.00,41.00)
\bezier{208}(79.00,48.00)(72.70,68.33)(79.00,99.00)
\bezier{36}(79.00,99.00)(80.00,93.50)(80.00,89.00)
\bezier{116}(80.00,89.00)(78.00,78.50)(80.00,60.00)
\bezier{52}(80.00,60.00)(80.00,54.00)(79.00,48.00)
\bezier{96}(40.00,88.00)(42.00,72.00)(40.00,64.00)
\bezier{52}(41.00,101.00)(40.00,93.50)(40.00,88.00)
\bezier{48}(41.00,52.00)(40.00,60.00)(40.00,64.00)
\thinlines
\bezier{50}(41.00,101.00)(47.70,79.33)(41.00,52.00)
\bezier{16}(63.00,59.00)(65.67,61.33)(66.00,63.00)
\bezier{54}(66.00,63.00)(79.67,79.33)(63.00,107.00)
\bezier{108}(63.00,59.00)(53.56,71.11)(54.00,83.00)
\bezier{28}(54.00,83.00)(53.56,83.11)(55.00,90.00)
\bezier{20}(55.00,90.00)(56.89,96.89)(63.00,108.00)
\bezier{44}(63.00,59.00)(62.00,56.50)(61.00,48.00)
\bezier{23}(61.00,48.00)(60.50,37.50)(60.00,25.00)
\thicklines
\bezier{152}(41.00,52.00)(54.00,47.33)(60.00,25.00)
\bezier{156}(60.00,25.00)(65.00,48.00)(79.00,48.00)
\bezier{84}(57.00,41.00)(59.67,43.00)(60.00,25.00)
\bezier{148}(41.00,101.00)(56.00,105.00)(60.00,125.00)
\bezier{148}(60.00,125.00)(66.00,104.00)(79.00,99.00)
\bezier{132}(57.00,92.00)(59.67,103.33)(60.00,125.00)
\thinlines
\bezier{19}(60.00,125.00)(62.00,107.33)(63.00,108.00)
\put(25.00,100.00){\vector(4,-1){31.00}}
\put(25.00,101.00){\vector(1,0){15.50}}
\put(24.00,101.00){\makebox(0,0)[rc]{$P$}}
\put(96.00,99.00){\vector(-1,0){16.50}}
\bezier{76}(96.00,101.00)(88.00,107.00)(80.00,109.00)
\put(63.00,109.00){\vector(-4,-1){0.2}}
\bezier{68}(80.00,109.00)(72.67,111.00)(63.00,109.00)
\put(97.00,100.00){\makebox(0,0)[lc]{$P$}}
\bezier{72}(53.30,87.00)(49.50,77.73)(49.50,70.00)
\bezier{64}(49.50,70.00)(49.48,62.88)(53.27,55.00)
\bezier{52}(53.27,55.00)(56.30,48.79)(58.12,43.03)
\end{picture}

\smallskip
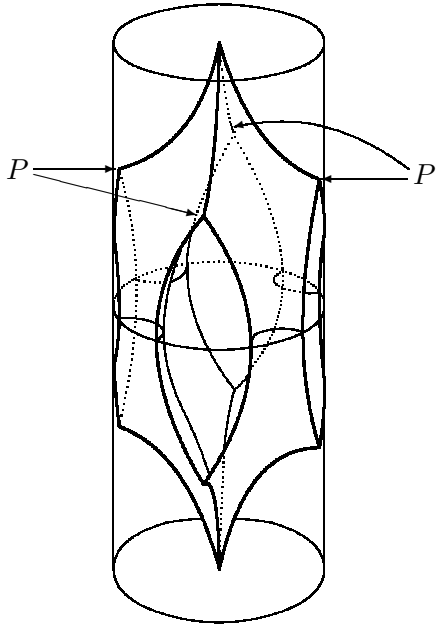
\noindent {\small Fig.~4: The fundamental region of
the new black hole. The points
labelled by $P$ are really the same point, and the event horizon
is the backwards light cone of this point. (In stereographic
coordinates centered on the time-symmetric surface the topology of
this figure would be the same, but all surfaces, including the
limiting cylinder, would become hyperboloids. For a stereographic
cross section see Fig.~10.)}
}

\vspace{8ex}
\noindent
{\large IV. Spatial Topology.}

\vspace{3ex}

\noindent In this section we discuss the various
topologies that occur as solutions of Einstein's equations
with a negative cosmological constant, and find their global
degrees of freedom. We confine attention to
time-symmetric geometries with a single asymptotically de Sitter region.
The extension of the analysis to cases with several asymptotic
regions is quite straightforward.

We will consider the initial geometry on a surface of time symmetry.
The only condition on this 2-dimensional
spacelike geometry is the Einstein constraint, which in the case under
consideration (time symmetry means vanishing extrinsic curvature)
demands that the intrinsic geometry have constant curvature.
If space is closed this means that all topologies with genus $g > 1$
are admitted, and no others. But we will consider the non-compact
case.

We have found two ways of constructing such 2-dimensional spaces
S useful. We made use of the first already in section III, where
the initial data slice appeared as the quotient by a suitable
group of isometries\footnote{As remarked before, in this paper
we exclude groups with fixed points on the initial surface, which
would give rise to particle-like singularities. This is to be contrasted
with the spacetime isometry group, where we do want fixed points in
order to generate black holes.}
of the universal cover, the simply connected space H$^2$ of constant negative
curvature, conveniently represented as the Poincar\'e disk.

\vspace{-0.5cm}
\noindent
\parbox{\textwidth}{
\unitlength 0.71mm
\linethickness{0.4pt}
\begin{picture}(192.77,114.77)(8,30)
\bezier{11}(35.23,100.00)(35.23,105.54)(39.54,110.46)
\bezier{11}(39.54,110.46)(44.46,114.77)(50.00,114.77)
\bezier{11}(64.77,100.00)(64.77,105.54)(60.46,110.46)
\bezier{11}(60.46,110.46)(55.54,114.77)(50.00,114.77)
\bezier{11}(35.23,100.00)(35.23,94.46)(39.54,89.54)
\bezier{11}(39.54,89.54)(44.46,85.23)(50.00,85.23)
\bezier{11}(64.77,100.00)(64.77,94.46)(60.46,89.54)
\bezier{11}(60.46,89.54)(55.54,85.23)(50.00,85.23)
\bezier{40}(42.55,112.56)(45.36,108.58)(50.00,108.58)
\bezier{40}(57.45,112.56)(54.64,108.58)(50.00,108.58)
\bezier{40}(62.56,107.45)(58.58,104.64)(58.58,100.00)
\bezier{40}(62.56,92.55)(58.58,95.36)(58.58,100.00)
\bezier{40}(57.45,87.44)(54.64,91.42)(50.00,91.42)
\bezier{40}(42.55,87.44)(45.36,91.42)(50.00,91.42)
\bezier{40}(37.44,107.45)(41.42,104.64)(41.42,100.00)
\put(41.33,99.11){\vector(0,1){0.89}}
\put(41.33,100.00){\vector(0,1){0.89}}
\put(58.67,99.11){\vector(0,1){0.89}}
\put(58.67,100.00){\vector(0,1){0.89}}
\put(49.56,91.33){\vector(1,0){0.89}}
\put(49.56,108.67){\vector(1,0){0.89}}
\thicklines
\bezier{16}(50.00,91.56)(50.00,100.00)(50.00,108.44)
\thinlines
\bezier{11}(85.23,100.00)(85.23,105.54)(89.54,110.46)
\bezier{11}(89.54,110.46)(94.46,114.77)(100.00,114.77)
\bezier{11}(114.77,100.00)(114.77,105.54)(110.46,110.46)
\bezier{11}(110.46,110.46)(105.54,114.77)(100.00,114.77)
\bezier{11}(85.23,100.00)(85.23,94.46)(89.54,89.54)
\bezier{11}(89.54,89.54)(94.46,85.23)(100.00,85.23)
\bezier{11}(114.77,100.00)(114.77,94.46)(110.46,89.54)
\bezier{11}(110.46,89.54)(105.54,85.23)(100.00,85.23)
\bezier{40}(92.55,112.56)(95.36,108.58)(100.00,108.58)
\bezier{40}(92.55,87.44)(95.36,91.42)(100.00,91.42)
\bezier{40}(87.44,92.55)(91.42,95.36)(91.42,100.00)
\bezier{40}(87.44,107.45)(91.42,104.64)(91.42,100.00)
\put(91.33,99.11){\vector(0,1){0.89}}
\put(91.33,100.00){\vector(0,1){0.89}}
\put(100.00,91.56){\line(0,1){16.89}}
\put(100.00,98.67){\vector(0,1){0.44}}
\put(100.00,99.11){\vector(0,1){0.89}}
\put(100.00,100.00){\vector(0,1){0.89}}
\bezier{11}(164.77,100.00)(164.77,105.54)(160.46,110.46)
\bezier{11}(160.46,110.46)(155.54,114.77)(150.00,114.77)
\bezier{11}(135.23,100.00)(135.23,105.54)(139.54,110.46)
\bezier{11}(139.54,110.46)(144.46,114.77)(150.00,114.77)
\bezier{11}(164.77,100.00)(164.77,94.46)(160.46,89.54)
\bezier{11}(160.46,89.54)(155.54,85.23)(150.00,85.23)
\bezier{11}(135.23,100.00)(135.23,94.46)(139.54,89.54)
\bezier{11}(139.54,89.54)(144.46,85.23)(150.00,85.23)
\bezier{40}(157.45,112.56)(154.64,108.58)(150.00,108.58)
\bezier{40}(157.45,87.44)(154.64,91.42)(150.00,91.42)
\bezier{40}(162.56,92.55)(158.58,95.36)(158.58,100.00)
\bezier{40}(162.56,107.45)(158.58,104.64)(158.58,100.00)
\put(158.67,99.11){\vector(0,1){0.89}}
\put(158.67,100.00){\vector(0,1){0.89}}
\put(150.00,91.56){\line(0,1){16.89}}
\put(150.00,98.67){\vector(0,1){0.44}}
\put(150.00,99.11){\vector(0,1){0.89}}
\put(150.00,100.00){\vector(0,1){0.89}}
\put(75.00,100.00){\makebox(0,0)[cc]{\large =}}
\put(125.00,100.00){\makebox(0,0)[cc]{\large +}}
\put(94.78,110.00){\vector(3,-2){1.33}}
\put(94.78,90.00){\vector(3,2){1.33}}
\put(155.22,110.00){\vector(-3,-2){1.33}}
\put(155.22,90.00){\vector(-3,2){1.33}}
\bezier{11}(11.23,60.00)(11.23,54.46)(15.54,49.54)
\bezier{11}(15.54,49.54)(20.46,45.23)(26.00,45.23)
\bezier{11}(40.77,60.00)(40.77,54.46)(36.46,49.54)
\bezier{11}(36.46,49.54)(31.54,45.23)(26.00,45.23)
\bezier{11}(11.23,60.00)(11.23,65.54)(15.54,70.46)
\bezier{11}(15.54,70.46)(20.46,74.77)(26.00,74.77)
\bezier{11}(40.77,60.00)(40.77,65.54)(36.46,70.46)
\bezier{11}(36.46,70.46)(31.54,74.77)(26.00,74.77)
\put(26.00,45.11){\line(0,1){29.78}}
\put(26.00,58.67){\vector(0,1){0.44}}
\put(26.00,59.11){\vector(0,1){1.11}}
\bezier{16}(35.47,64.22)(34.67,62.36)(34.53,59.96)
\bezier{56}(32.80,73.30)(29.20,67.50)(35.47,64.22)
\bezier{16}(35.47,64.22)(34.53,62.36)(34.53,60.00)
\bezier{16}(35.47,55.78)(34.67,57.64)(34.53,60.04)
\bezier{56}(32.80,46.70)(29.20,52.50)(35.47,55.78)
\put(31.67,68.22){\vector(1,-3){0.44}}
\put(34.83,59.39){\vector(0,1){0.33}}
\put(34.83,59.72){\vector(0,1){0.67}}
\put(34.83,60.39){\vector(0,1){0.67}}
\put(31.67,51.78){\vector(1,3){0.44}}
\bezier{11}(90.77,60.00)(90.77,54.46)(86.46,49.54)
\bezier{11}(86.46,49.54)(81.54,45.23)(76.00,45.23)
\bezier{11}(61.23,60.00)(61.23,54.46)(65.54,49.54)
\bezier{11}(65.54,49.54)(70.46,45.23)(76.00,45.23)
\bezier{11}(90.77,60.00)(90.77,65.54)(86.46,70.46)
\bezier{11}(86.46,70.46)(81.54,74.77)(76.00,74.77)
\bezier{11}(61.23,60.00)(61.23,65.54)(65.54,70.46)
\bezier{11}(65.54,70.46)(70.46,74.77)(76.00,74.77)
\put(76.00,45.11){\line(0,1){29.78}}
\put(76.00,58.67){\vector(0,1){0.44}}
\put(76.00,59.11){\vector(0,1){1.11}}
\bezier{16}(66.53,64.22)(67.33,62.36)(67.47,59.96)
\bezier{56}(69.20,73.30)(72.80,67.50)(66.53,64.22)
\bezier{16}(66.53,64.22)(67.47,62.36)(67.47,60.00)
\bezier{16}(66.53,55.78)(67.33,57.64)(67.47,60.04)
\bezier{56}(69.20,46.70)(72.80,52.50)(66.53,55.78)
\put(70.33,68.22){\vector(-1,-3){0.44}}
\put(67.50,59.39){\vector(0,1){0.33}}
\put(67.50,59.72){\vector(0,1){0.67}}
\put(67.50,60.39){\vector(0,1){0.67}}
\put(70.33,51.78){\vector(-1,3){0.44}}
\put(51.00,60.00){\makebox(0,0)[cc]{\large +}}
\put(175.00,98.00){\makebox(0,0)[cc]{\large =}}
\bezier{11}(142.77,60.00)(142.77,54.46)(138.46,49.54)
\bezier{11}(138.46,49.54)(133.54,45.23)(128.00,45.23)
\bezier{11}(113.23,60.00)(113.23,54.46)(117.54,49.54)
\bezier{11}(117.54,49.54)(122.46,45.23)(128.00,45.23)
\bezier{11}(142.77,60.00)(142.77,65.54)(138.46,70.46)
\bezier{11}(138.46,70.46)(133.54,74.77)(128.00,74.77)
\bezier{11}(113.23,60.00)(113.23,65.54)(117.54,70.46)
\bezier{11}(117.54,70.46)(122.46,74.77)(128.00,74.77)
\bezier{16}(118.53,64.22)(119.33,62.36)(119.47,59.96)
\bezier{56}(121.20,73.30)(124.80,67.50)(118.53,64.22)
\bezier{16}(118.53,64.22)(119.47,62.36)(119.47,60.00)
\bezier{16}(118.53,55.78)(119.33,57.64)(119.47,60.04)
\bezier{56}(121.20,46.70)(124.80,52.50)(118.53,55.78)
\put(122.33,68.22){\vector(-1,-3){0.44}}
\put(119.50,59.3){\vector(0,1){0.33}}
\put(119.50,60.0){\vector(0,1){0.67}}
\put(119.50,61.0){\vector(0,1){0.67}}
\put(122.33,51.78){\vector(-1,3){0.44}}
\bezier{16}(137.47,64.22)(136.67,62.36)(136.53,59.96)
\bezier{56}(134.80,73.30)(131.20,67.50)(137.47,64.22)
\bezier{16}(137.47,64.22)(136.53,62.36)(136.53,60.00)
\bezier{16}(137.47,55.78)(136.67,57.64)(136.53,60.04)
\bezier{56}(134.80,46.70)(131.20,52.50)(137.47,55.78)
\put(133.67,68.22){\vector(1,-3){0.44}}
\put(136.50,59.3){\vector(0,1){0.33}}
\put(136.50,60.0){\vector(0,1){0.67}}
\put(136.50,61.0){\vector(0,1){0.67}}
\put(133.67,51.78){\vector(1,3){0.44}}
\put(134.00,67.55){\vector(1,-3){0.44}}
\put(134.00,52.45){\vector(1,3){0.44}}
\thicklines
\bezier{12}(119.50,60.00)(127.83,60.00)(136.33,60.00)
\thinlines
\put(103.00,60.00){\makebox(0,0)[cc]{\large =}}
\bezier{11}(192.77,60.00)(192.77,54.46)(188.46,49.54)
\bezier{11}(188.46,49.54)(183.54,45.23)(178.00,45.23)
\bezier{11}(163.23,60.00)(163.23,54.46)(167.54,49.54)
\bezier{11}(167.54,49.54)(172.46,45.23)(178.00,45.23)
\bezier{11}(192.77,60.00)(192.77,65.54)(188.46,70.46)
\bezier{11}(188.46,70.46)(183.54,74.77)(178.00,74.77)
\bezier{11}(163.23,60.00)(163.23,65.54)(167.54,70.46)
\bezier{11}(167.54,70.46)(172.46,74.77)(178.00,74.77)
\bezier{16}(168.53,64.22)(169.33,62.36)(169.47,59.96)
\bezier{56}(171.20,73.30)(174.80,67.50)(168.53,64.22)
\bezier{16}(168.53,64.22)(169.47,62.36)(169.47,60.00)
\put(172.33,68.22){\vector(-1,-3){0.44}}
\bezier{16}(187.47,64.22)(186.67,62.36)(186.53,59.96)
\bezier{56}(184.80,73.30)(181.20,67.50)(187.47,64.22)
\bezier{16}(187.47,64.22)(186.53,62.36)(186.53,60.00)
\put(183.67,68.22){\vector(1,-3){0.44}}
\put(184.00,67.55){\vector(1,-3){0.44}}
\put(186.22,60.00){\line(-1,0){16.44}}
\put(176.78,60.00){\vector(1,0){0.44}}
\put(177.22,60.00){\vector(1,0){1.11}}
\put(178.33,60.00){\vector(1,0){1.11}}
\put(179.44,60.00){\vector(1,0){1.11}}
\put(155.00,60.00){\makebox(0,0)[cc]{\large = \ \Large 2$\times$}}
\bezier{52}(171.96,67.47)(177.96,64.80)(183.96,67.47)
\put(178.00,71.00){\makebox(0,0)[cc]{$\cal E$}}
\put(178.00,63.00){\makebox(0,0)[cc]{$\cal T$}}
\put(169.56,61.33){\vector(-1,4){0.17}}
\put(169.39,62.00){\vector(-1,4){0.17}}
\put(169.22,62.67){\vector(-1,4){0.17}}
\put(186.44,61.33){\vector(1,4){0.17}}
\put(186.61,62.00){\vector(1,4){0.17}}
\put(186.78,62.67){\vector(1,4){0.17}}
\put(42.00,100.00){\circle*{0.00}}
\put(50.00,92.00){\circle*{0.00}}
\bezier{40}(37.44,92.55)(41.42,95.36)(41.42,100.00)
\end{picture}

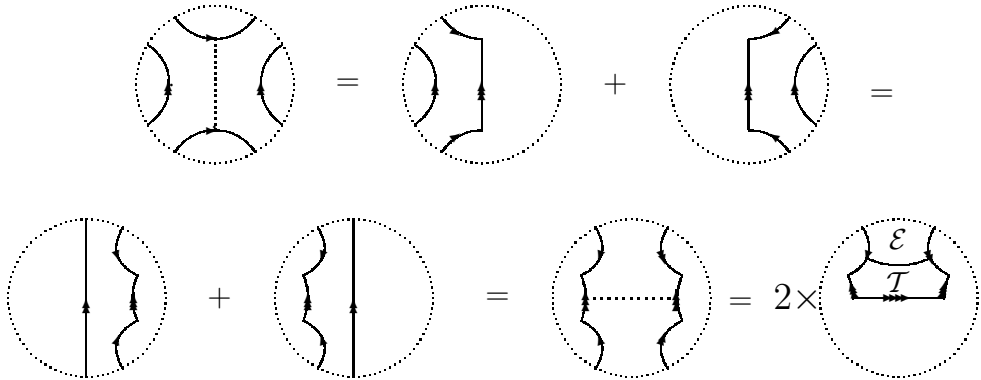
\noindent {\small Fig.~5: Different but equivalent ways to construct
initial data of a black hole of the wormhole type. We begin with a
fundamental region of the kind that was used in the previous section,
having one asymptotic region only (although this is not immediately
apparent from the picture). Then we cut the disk in two, apply a
global isometry to the two pieces, and glue them back together in
a new way. Finally we cut the new disk in two to obtain two disks,
which we refer to a ``doubling'' of the initial data. The fundamental
region on the new pair of disks naturally splits in two pieces, a
hyperbolic hexagon $\cal T$ representing the interior of the black
hole together with an exterior region denoted $\cal E$. Gluing the
hyperbolic hexagons together we obtain the ``pair of pants'' used
in Figure 6.}
}
\vspace{5mm}

For each such isometry group one can find a fundamental
region that yields S when its boundaries are identified according to the
isometries. It is always possible to choose this region to have
geodesic boundaries.  The first of the disks shown in Figure 5
illustrates a fundamental region for a black hole with a
single exterior and a toroidal interior, which is the type of
black hole of primary interest in the following sections.
Boundaries bearing an equal number of arrowheads are to be
identified with each other.
An advantage of this representation is that only a
single region is needed; a disadvantage is that for more
complicated topologies the identifications are not easy to
visualize, nor is it apparent how to count the parameters that
specify the geometry (as a matter of fact there are three
parameters, the geodesic distance between the boundaries and
an additional angle).

The second construction builds S combinatorially out of simple blocks. The
boundaries of these building blocks are chosen to be geodesic, so the only
smoothness condition (which determines whether two blocks fit together)
is that the lengths of the boundaries agree. The blocks can be easily
visualized by ``doubling'' a region bounded by 3 resp.\ 6 geodesics.
The last part of Figure 5 shows shows such a doubling, where unmatched
boundaries are identified between the first and second copy of the figure.
(Note that mathematicians use the word ``doubling'' in a different
way \cite{Mathematics}).
When the pair of disk regions is glued together in this way we obtain
a three dimensional picture of the initial data as shown in Figure 6a. A
minimal geodesic divides it into a region called ``trousers'' $\cal T$ and
an exterior region $\cal E$.

\vspace{0.5cm}
\noindent
\parbox{\textwidth}{
\unitlength 1.00mm
\linethickness{0.4pt}
\begin{picture}(117.00,70.00)(0,5)
\put(30.00,30.00){\circle{10.00}}
\put(27.00,46.00){\line(-1,0){8.00}}
\put(19.00,46.00){\line(-1,1){6.00}}
\thicklines
\put(13.00,52.00){\line(1,0){17.00}}
\thinlines
\put(33.00,46.00){\line(1,0){8.00}}
\put(41.00,46.00){\line(1,1){6.00}}
\thicklines
\put(47.00,52.00){\line(-1,0){17.00}}
\put(30.00,49.00){\makebox(0,0)[cc]{$\cal E$}}
\bezier{32}(19.78,30.00)(19.78,34.14)(22.71,37.07)
\bezier{32}(39.78,30.00)(39.78,34.14)(36.85,37.07)
\bezier{32}(19.78,30.00)(19.78,25.86)(22.71,22.93)
\bezier{32}(22.71,22.93)(25.64,20.00)(29.78,20.00)
\bezier{32}(39.78,30.00)(39.78,25.86)(36.85,22.93)
\bezier{32}(36.85,22.93)(33.92,20.00)(29.78,20.00)
\bezier{8}(27.13,44.00)(27.13,44.53)(27.95,44.91)
\bezier{8}(27.95,44.91)(28.77,45.29)(29.94,45.29)
\bezier{8}(32.75,44.00)(32.75,44.53)(31.93,44.91)
\bezier{8}(31.93,44.91)(31.10,45.29)(29.94,45.29)
\thinlines
\bezier{8}(27.13,44.00)(27.13,43.47)(27.95,43.09)
\bezier{8}(27.95,43.09)(28.77,42.71)(29.94,42.71)
\bezier{8}(32.75,44.00)(32.75,43.47)(31.93,43.09)
\bezier{8}(31.93,43.09)(31.10,42.71)(29.94,42.71)
\bezier{24}(32.75,44.00)(32.75,47.00)(37.00,48.40)
\bezier{24}(27.21,44.00)(27.21,47.00)(22.96,48.40)
\thicklines
\bezier{8}(27.13,40.40)(27.13,40.93)(27.95,41.31)
\bezier{8}(27.95,41.31)(28.77,41.69)(29.94,41.69)
\bezier{8}(32.75,40.40)(32.75,40.93)(31.93,41.31)
\bezier{8}(31.93,41.31)(31.10,41.69)(29.94,41.69)
\bezier{16}(36.84,37.07)(35.64,38.40)(33.38,39.60)
\bezier{4}(33.38,39.60)(32.98,39.87)(32.71,40.53)
\bezier{16}(22.84,37.07)(24.04,38.40)(26.31,39.60)
\bezier{4}(26.31,39.60)(26.71,39.87)(26.98,40.53)
\thinlines
\bezier{16}(30.18,24.93)(31.24,25.07)(31.38,22.40)
\bezier{12}(31.38,22.40)(30.98,20.53)(30.04,20.13)
\put(97.00,64.00){\line(-1,0){8.00}}
\put(89.00,64.00){\line(-1,1){6.00}}
\thicklines
\put(83.00,70.00){\line(1,0){17.00}}
\put(117.00,70.00){\line(-1,0){17.00}}
\thinlines
\put(103.00,64.00){\line(1,0){8.00}}
\put(111.00,64.00){\line(1,1){6.00}}
\put(100.00,67.00){\makebox(0,0)[cc]{$\cal E$}}
\thicklines
\bezier{8}(97.13,62.00)(97.13,62.53)(97.95,62.91)
\bezier{8}(97.95,62.91)(98.77,63.29)(99.94,63.29)
\bezier{8}(102.75,62.00)(102.75,62.53)(101.93,62.91)
\bezier{8}(101.93,62.91)(101.10,63.29)(99.94,63.29)
\thinlines
\bezier{8}(97.13,62.00)(97.13,61.47)(97.95,61.09)
\bezier{8}(97.95,61.09)(98.77,60.71)(99.94,60.71)
\bezier{8}(102.75,62.00)(102.75,61.47)(101.93,61.09)
\bezier{8}(101.93,61.09)(101.10,60.71)(99.94,60.71)
\bezier{24}(102.75,62.00)(102.75,65.00)(107.00,66.40)
\bezier{24}(97.21,62.00)(97.21,65.00)(92.96,66.40)
\put(100.00,51.00){\circle{8.00}}
\thicklines
\bezier{8}(97.13,59.00)(97.13,59.53)(97.95,59.91)
\bezier{8}(97.95,59.91)(98.77,60.29)(99.94,60.29)
\bezier{8}(102.75,59.00)(102.75,59.53)(101.93,59.91)
\bezier{8}(101.93,59.91)(101.10,60.29)(99.94,60.29)
\bezier{8}(96.84,42.05)(96.84,42.64)(97.74,43.05)
\bezier{8}(97.74,43.05)(98.64,43.47)(99.93,43.47)
\bezier{8}(103.03,42.05)(103.03,42.64)(102.13,43.05)
\bezier{8}(102.13,43.05)(101.21,43.47)(99.93,43.47)
\thinlines
\bezier{8}(96.84,42.05)(96.84,41.47)(97.74,41.05)
\bezier{8}(97.74,41.05)(98.64,40.63)(99.93,40.63)
\bezier{8}(103.03,42.05)(103.03,41.47)(102.13,41.05)
\bezier{8}(102.13,41.05)(101.21,40.63)(99.93,40.63)
\thicklines
\bezier{12}(102.81,58.97)(102.81,57.90)(104.09,57.44)
\bezier{36}(104.09,57.44)(107.94,55.38)(107.94,51.03)
\bezier{36}(104.09,44.40)(107.94,46.67)(107.94,51.03)
\bezier{20}(104.09,44.40)(103.00,44.00)(103.00,42.26)
\bezier{12}(97.19,58.97)(97.19,57.90)(95.91,57.44)
\bezier{36}(95.91,57.44)(92.06,55.38)(92.06,51.03)
\bezier{36}(95.91,44.40)(92.06,46.67)(92.06,51.03)
\bezier{20}(95.91,44.40)(97.00,44.00)(97.00,42.26)
\thinlines
\put(100.00,28.00){\circle{10.00}}
\thicklines
\bezier{32}(89.78,28.00)(89.78,32.14)(92.71,35.07)
\bezier{32}(109.78,28.00)(109.78,32.14)(106.85,35.07)
\bezier{32}(89.78,28.00)(89.78,23.86)(92.71,20.93)
\bezier{32}(92.71,20.93)(95.64,18.00)(99.78,18.00)
\bezier{32}(109.78,28.00)(109.78,23.86)(106.85,20.93)
\bezier{32}(106.85,20.93)(103.92,18.00)(99.78,18.00)
\bezier{8}(97.13,38.40)(97.13,38.93)(97.95,39.31)
\bezier{8}(97.95,39.31)(98.77,39.69)(99.94,39.69)
\bezier{8}(102.75,38.40)(102.75,38.93)(101.93,39.31)
\bezier{8}(101.93,39.31)(101.10,39.69)(99.94,39.69)
\bezier{16}(106.84,35.07)(105.64,36.40)(103.38,37.60)
\bezier{4}(103.38,37.60)(102.98,37.87)(102.71,38.53)
\bezier{16}(92.84,35.07)(94.04,36.40)(96.31,37.60)
\bezier{4}(96.31,37.60)(96.71,37.87)(96.98,38.53)
\thinlines
\bezier{16}(100.18,22.93)(101.24,23.07)(101.38,20.40)
\bezier{12}(101.38,20.40)(100.98,18.53)(100.04,18.13)
\bezier{20}(92.14,50.97)(94.02,52.0)(95.91,50.97)
\bezier{20}(107.79,50.97)(105.91,52.0)(104.02,50.97)
\put(100.00,57.00){\makebox(0,0)[cc]{$\cal T$}}
\put(100.00,44.00){\makebox(0,0)[cb]{$\cal T$}}
\put(100.00,36.00){\makebox(0,0)[cc]{$\cal T$}}
\put(30.00,38.00){\makebox(0,0)[cc]{$\cal T$}}
\put(51.00,42.00){\vector(-1,0){16.00}}
\put(52.00,42.00){\makebox(0,0)[lc]{Minimal circles}}
\put(76.00,43.00){\vector(2,1){15.00}}
\put(77.75,41.5){\vector(1,0){18.00}}
\put(32.00,19.00){\vector(-1,1){0.2}}
\bezier{230}(52.00,40.00)(40.00,10.51)(32.00,19.00)
\put(30.00,10.00){\makebox(0,0)[cc]{a}}
\put(100.00,10.00){\makebox(0,0)[cc]{b}}
\end{picture}

\vspace{5mm}

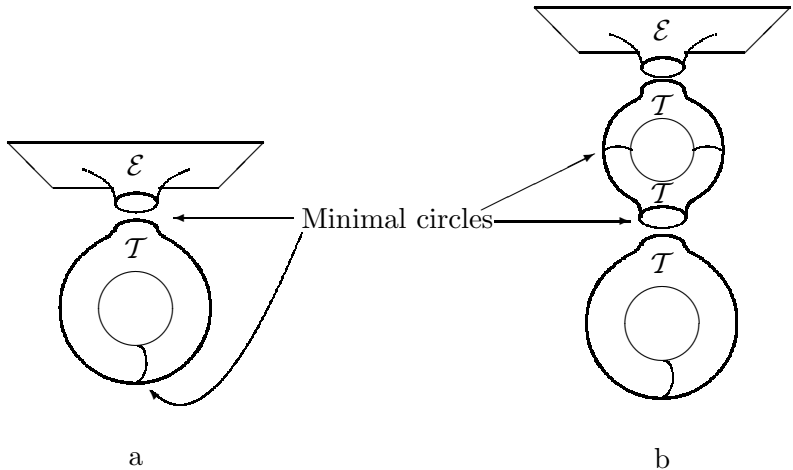
\noindent {\small Fig.~6: Construction of the class of spaces
considered here by sewing together a set of trousers and one
asymptotic region (which looks asymptotically flat in the picture,
but which does in fact have constant negative curvature
everywhere, just like the trousers).}
}
\vspace{5mm}

Regions of these two types can also be glued together to form more
complicated topologies (Figure 6b).
By an (orientable) 2-surface of genus $g$ with one asymptotic region we mean
a topology of the type
${\rm I\!R}^2$$\#{\rm T}^2\#$$\ldots$$(g\; {\rm factors)}$$\ldots$$\#{\rm
T}^2$,
 that is a plane with $g$ handles attached.
For any noncontractible circle in a space of this topology
there is a homotopic minimal circle. We can choose $3g-1$ such circles that
divide the surface into an exterior $\cal E$ and $2g-1$ trousers $\cal T$.
Figure 6a shows how this is done for the black hole with a
toroidal interior, ${\rm I\!R}^2\#{\rm T}^2$, and Figure 6b shows
how further handles are inserted.

In this description we can choose as the parameters the horizon length of
the exterior, the length around each opening of the trousers, and the
amount of twist by which two openings are turned before identification.
Thus in Fig 6a the trouser ``legs'' that are identified have to have the
same circumference, and the trouser ``waist'' must have the same circumference
as the exterior horizon, so there are two length parameters. In addition there
is one twist parameter at the legs. (The twist at the waist does not change
the geometry, because the exterior is rotationally symmetric.) Thus we find
3 parameters determining this geometry, as above.
It follows from figure~6b that adding a handle yields
3 more lengths and 3 more twists, so the genus $g$
geometry is characterized by $6g-3$ parameters.
These parameters span a $(6g - 3)$-dimensional space known as
Teichm\"{u}ller space; it is the space
of two-geometries modulo the connected component of the diffeomorphism
group and modulo conformal transformations. (This is a well known
mathematical construction \cite{Mathematics}. In our case the conformal
factor is fixed by the requirement of constant curvature. The length
and twist parameters
are known as Fenchel-Nielsen coordinates on Teichm\"{u}ller space.)

The moduli space or ``superspace'' (in the sense of Wheeler
and Fischer \cite{WF}) of these 2-geometries
is also $(6g-3)$-dimensional, but due to invariances under ``large''
diffeomorphisms it has a non-trivial topology. (For example, a twist by
$2\pi$ leaves the geometry unchanged; two wormholes that
``look'' different may nevertheless have the same intrinsic
geometry.) Thus the superspace of our black hole initial
states has its usual structure
of a stratified manifold, with the strata representing geometries of
greater symmetry; it is more difficult to understand in detail
than Teichm\"{u}ller space.

\vspace{5ex}
\noindent
{\large V. Time Development of the Wormholes.}

\vspace{3ex}

\noindent The simplest wormhole is evidently the toroidal one,
so let us focus on this case first. We also choose the fundamental
region to be as symmetric as possible, so that the discrete isometry
group ${\Gamma}$ that it defines is generated by the isometries
$a = e^{{\gamma}J_{TX}}$ and $b = e^{{\gamma}J_{TY}}$, with the same
real number ${\gamma}$ in both cases.
(That is, we concentrate on a one parameter family of toroidal
wormholes, rather than on the three parameter family that we have
shown to exist.)
Here $J_{TX}$ and $J_{TY}$ are the Killing vectors

\begin{equation} J_{TX} = X\partial_T + T\partial_X \hspace{1cm}
J_{TY} = Y\partial_T + T\partial_Y \ . \end{equation}

\noindent They clearly preserve the surface $U = 0$, and we assume
that ${\gamma}$ is large enough so that an asymptotic region exists.
Next we extend the action of the isometries
to anti-de Sitter space,
and we choose  the fundamental
region as depicted in figure 4 above (using sausage coordinates).
As noted, the folds in the roof of the ``tent'' define lines of
fixed points of some of the isometries; we pushed the fixed points
away from the initial disk by opening an asymptotic region, but
with the passage of time they are coming back into our spacetime.
However, since the folds cannot be seen from infinity such
singularities are allowed. It is
also worth observing that our spacetime has no global
Killing vector (this is evident since such a Killing vector would
have to commute with both $J_{TX}$ and $J_{TY}$, which is
impossible); it may of course still have discrete symmetries.

The event horizon is defined as a surface that bounds a region
of spacetime which cannot be seen from \scri, which in our picture
consists of the openings of the ``tent''.
It can actually be a little tricky to find the event horizon from a given
choice of fundamental region. However, our choice does not belong
to the tricky ones, since the identifications are such that they
can never cause a curve to turn back in ``sausage time'' $t$. It
is therefore evident that the event horizon
is given by the backwards light cone of the last point on \scri \
(in Figure 4 this appears as four points to be identified,
marked $P$).

For a first study of the exterior of this black hole, let us return
to the initial data point of view, and more particularly to the
series of cuttings and gluings that turned our original fundamental
region into two hyperbolic hexagons. This time we do not glue
these together into a pair of pants, but into a single region in
the Poincar\'{e} disk as in Figure 7.
It is now manifest that
there is one asymptotic region only; moreover the event horizon
is easily located since---as explained in section III---it must
be the unique geodesic that is to
be found among the flowlines of the transvection that identifies
the two external boundaries of the fundamental region.

\vspace{1cm}
\noindent
\parbox{\textwidth}{
\unitlength 0.90mm
\linethickness{0.4pt}
\begin{picture}(149.77,34.83)(8,0)
\bezier{11}(49.77,20.00)(49.77,14.46)(45.46,9.54)
\bezier{11}(45.46,9.54)(40.54,5.23)(35.00,5.23)
\bezier{11}(20.23,20.00)(20.23,14.46)(24.54,9.54)
\bezier{11}(24.54,9.54)(29.46,5.23)(35.00,5.23)
\bezier{11}(49.77,20.00)(49.77,25.54)(45.46,30.46)
\bezier{11}(45.46,30.46)(40.54,34.77)(35.00,34.77)
\bezier{11}(20.23,20.00)(20.23,25.54)(24.54,30.46)
\bezier{11}(24.54,30.46)(29.46,34.77)(35.00,34.77)
\bezier{16}(25.53,24.22)(26.33,22.36)(26.47,19.96)
\bezier{56}(28.20,33.30)(31.80,27.50)(25.53,24.22)
\bezier{16}(25.53,24.22)(26.47,22.36)(26.47,20.00)
\put(29.33,28.22){\vector(-1,-3){0.44}}
\bezier{16}(44.47,24.22)(43.67,22.36)(43.53,19.96)
\bezier{56}(41.80,33.30)(38.20,27.50)(44.47,24.22)
\bezier{16}(44.47,24.22)(43.53,22.36)(43.53,20.00)
\put(40.67,28.22){\vector(1,-3){0.44}}
\put(41.00,27.55){\vector(1,-3){0.44}}
\put(43.22,20.00){\line(-1,0){16.44}}
\put(33.78,20.00){\vector(1,0){0.44}}
\put(34.22,20.00){\vector(1,0){1.11}}
\put(35.33,20.00){\vector(1,0){1.11}}
\put(36.44,20.00){\vector(1,0){1.11}}
\put(15.00,20.00){\makebox(0,0)[cc]{\Large 2$\times$}}
\bezier{52}(28.96,27.47)(34.96,24.80)(40.96,27.47)
\put(35.00,31.00){\makebox(0,0)[cc]{$\cal E$}}
\put(35.00,23.00){\makebox(0,0)[cc]{$\cal T$}}
\put(26.56,21.33){\vector(-1,4){0.17}}
\put(26.39,22.00){\vector(-1,4){0.17}}
\put(26.22,22.67){\vector(-1,4){0.17}}
\put(43.44,21.33){\vector(1,4){0.17}}
\put(43.61,22.00){\vector(1,4){0.17}}
\put(43.78,22.67){\vector(1,4){0.17}}
\bezier{11}(99.77,20.00)(99.77,14.46)(95.46,9.54)
\bezier{11}(95.46,9.54)(90.54,5.23)(85.00,5.23)
\bezier{11}(70.23,20.00)(70.23,14.46)(74.54,9.54)
\bezier{11}(74.54,9.54)(79.46,5.23)(85.00,5.23)
\bezier{11}(99.77,20.00)(99.77,25.54)(95.46,30.46)
\bezier{11}(95.46,30.46)(90.54,34.77)(85.00,34.77)
\bezier{11}(70.23,20.00)(70.23,25.54)(74.54,30.46)
\bezier{11}(74.54,30.46)(79.46,34.77)(85.00,34.77)
\put(62.00,20.00){\makebox(0,0)[cc]{\large = \ \Large 2$\times$}}
\put(85.00,34.83){\line(0,-1){22.67}}
\bezier{30}(85.00,12.33)(82.50,12.33)(80.00,10.83)
\bezier{34}(80.00,10.83)(78.83,13.50)(75.83,14.83)
\bezier{78}(75.83,14.83)(78.83,20.00)(75.83,25.17)
\bezier{82}(75.83,25.17)(81.50,27.50)(80.00,33.83)
\bezier{34}(79.67,28.17)(82.00,26.83)(85.00,26.83)
\put(82.67,30.83){\makebox(0,0)[cc]{$\cal E$}}
\put(81.00,20.00){\makebox(0,0)[cc]{$\cal T$}}
\put(85.00,21.00){\vector(0,-1){0.33}}
\put(85.00,20.40){\vector(0,-1){0.67}}
\put(79.67,28.17){\vector(-2,-3){0.33}}
\put(77.33,18.50){\vector(0,1){0.50}}
\put(77.33,19.00){\vector(0,1){1.00}}
\put(77.33,20.00){\vector(0,1){1.00}}
\put(76.83,14.33){\vector(1,-1){0.33}}
\put(77.17,14.00){\vector(1,-1){0.67}}
\put(77.83,13.33){\vector(1,-1){0.67}}
\put(78.50,12.67){\vector(1,-1){0.67}}
\put(81.33,11.67){\vector(3,1){0.50}}
\put(81.83,11.83){\vector(3,1){1.00}}
\put(82.83,12.17){\vector(1,0){0.83}}
\bezier{11}(149.77,20.00)(149.77,14.46)(145.46,9.54)
\bezier{11}(145.46,9.54)(140.54,5.23)(135.00,5.23)
\bezier{11}(120.23,20.00)(120.23,14.46)(124.54,9.54)
\bezier{11}(124.54,9.54)(129.46,5.23)(135.00,5.23)
\bezier{11}(149.77,20.00)(149.77,25.54)(145.46,30.46)
\bezier{11}(145.46,30.46)(140.54,34.77)(135.00,34.77)
\bezier{11}(120.23,20.00)(120.23,25.54)(124.54,30.46)
\bezier{11}(124.54,30.46)(129.46,34.77)(135.00,34.77)
\bezier{30}(135.00,12.33)(132.50,12.33)(130.00,10.83)
\bezier{34}(130.00,10.83)(128.83,13.50)(125.83,14.83)
\bezier{78}(125.83,14.83)(128.83,20.00)(125.83,25.17)
\bezier{82}(125.83,25.17)(131.50,27.50)(130.00,33.83)
\bezier{34}(129.67,28.17)(132.00,26.83)(135.00,26.83)
\put(129.67,28.17){\vector(-2,-3){0.33}}
\put(127.33,18.50){\vector(0,1){0.50}}
\put(127.33,19.00){\vector(0,1){1.00}}
\put(127.33,20.00){\vector(0,1){1.00}}
\put(126.83,14.33){\vector(1,-1){0.33}}
\put(127.17,14.00){\vector(1,-1){0.67}}
\put(127.83,13.33){\vector(1,-1){0.67}}
\put(128.50,12.67){\vector(1,-1){0.67}}
\bezier{30}(135.00,12.33)(137.50,12.33)(140.00,10.83)
\bezier{34}(140.00,10.83)(141.17,13.50)(144.17,14.83)
\bezier{78}(144.17,14.83)(141.17,20.00)(144.17,25.17)
\bezier{82}(144.17,25.17)(138.50,27.50)(140.00,33.83)
\bezier{34}(140.33,28.17)(138.00,26.83)(135.00,26.83)
\put(135.33,30.83){\makebox(0,0)[cc]{$\cal E$}}
\put(135.00,20.00){\makebox(0,0)[cc]{$\cal T$}}
\put(140.33,28.17){\vector(2,-3){0.33}}
\put(142.75,19.00){\vector(0,-1){0.50}}
\put(142.67,20.00){\vector(0,-1){0.50}}
\put(142.67,21.00){\vector(0,-1){0.50}}
\put(143.17,14.33){\vector(-1,-1){0.33}}
\put(142.83,14.00){\vector(-1,-1){0.67}}
\put(142.17,13.33){\vector(-1,-1){0.67}}
\put(141.50,12.67){\vector(-1,-1){0.67}}
\put(133.78,12.22){\vector(1,0){1.00}}
\put(134.78,12.22){\vector(1,0){1.00}}
\put(135.78,12.22){\vector(1,0){1.00}}
\put(110.00,20.00){\makebox(0,0)[cc]{\large =}}
\end{picture}

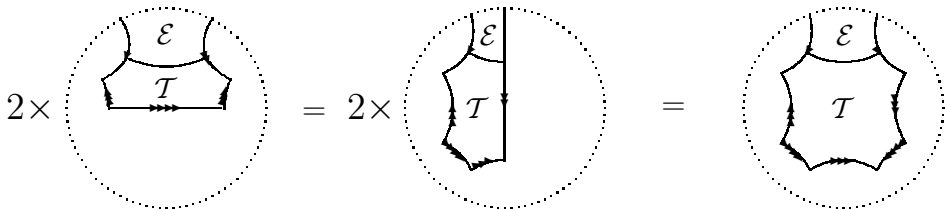
\noindent {\small Fig.~7: Making it manifest that our initial data leads
to one asymptotic region only.
We start with the last frame of Figure 5, apply a global isometry and glue the
two
parts together.
The line  dividing the regions $\cal E$ and
$\cal T$ is the event horizon at the moment of time symmetry.
\vspace{5mm}}
}

Now the point is that the region denoted $\cal E$
 is isometric to the exterior of the BTZ initial data. To the
future and past of this region there will be
a part of the exterior isometric to the exterior
of the BTZ black hole. It is a region of spacetime that
contains a static Killing vector, and to the future of the
moment of time symmetry it is bounded by the event horizon on
one side and by \scri \ on the other. We emphasize that this does
not mean that our wormhole is observationally indistinguishable
from the BTZ black hole. Indeed the BTZ exterior forms only
a subset of the exterior of the wormhole; to the past of the
moment of time symmetry the behavior of the event horizon
differs drastically between the two cases, as we will see. However,
the BTZ exterior agrees with the domain of outer communication
of the wormhole, which by definition is the set of points that
can be reached from \scri \ by causal curves in both the past and
the future direction.

Let us describe the construction analytically, using the embedding
coordinates of Eq (3) for this purpose. The fundamental region (depicted
in Figure 4) is bounded by the surfaces

\begin{equation} \frac{X}{T} = \pm \tanh{\frac{\gamma}{2}} \equiv
\pm {\alpha} \hspace{2cm} \frac{Y}{T} = \pm
\tanh{\frac{\gamma}{2}} \equiv \pm {\alpha} \ . \end{equation}

\noindent The real number ${\alpha}$ is as convenient a number
as any to parametrize our black hole. We observe that

\begin{equation} \frac{1}{\sqrt{2}} < {\alpha} < 1 \ ,
\end{equation}

\noindent and that the length of the event horizon grows with
${\alpha}$, so that a large value of the parameter
corresponds to a large mass for the black hole. The lower limit
on ${\alpha}$ guarantees that an asymptotic region exists---there
is no extremal black hole of this toroidal type that has
vanishing mass. The last point on
\scri \ occurs when the folds in
the roof of the tent reach the boundary of spacetime, which
happens for sausage time $t = t_P$ and stereographic time
${\tau} = {\tau}_P$, where

\begin{equation} \tan{t_P} = \sqrt{2{\alpha}^2 - 1}
= \frac{1}{{\tau}_P} \hspace{1cm}
{\rm hence} \hspace{1cm} 0 < t_P < \frac{\pi}{4} \ .
\end{equation}

\noindent Since it takes an amount ${\pi}/2$ of sausage time
for a radial light ray to go from the origin of adS to its boundary,
it is evident that the event horizon is born at the time
$t_{birth} = t_P - {\pi}/2 > - {\pi}/2$. But spacetime
itself is born at $t = - {\pi}/2$, i.e. before the event
horizon; this is not an eternal
black hole.

At late times the event horizon is the smooth backward light cone from
the last point on \scri . If such a point is given in stereographic
coordinates by $(x_P, y_P, {\tau}_P)$ then the backward light
cone is given in stereographic coordinates by

\begin{equation} (x - x_P)^2 + (y - y_P)^2 - ({\tau} - {\tau}_P)^2
= 0 \ . \end{equation}

\noindent When we use stereographic coordinates this appears
as a null cone with its vertex on the boundary of spacetime
similar to fig 1a,
but in anti-de Sitter space itself it is really a null plane
since its null generators have zero convergence and the vertex
is infinitely far away. In embedding coordinates the expression
becomes

\begin{equation} x_PX + y_PY - {\tau}_PT - U = 0 \ . \end{equation}

\noindent There will actually be four such surfaces in our
fundamental region, since the last point on \scri \ appears
as four points to be identified. Eventually the surfaces
will intersect, which leads to caustics in the horizon
(where new generators are added to it).
In order to understand the shape of the horizon
we use sausage coordinates and draw a sequence of equal-$t$ slices.
To locate the event horizon it is easiest to draw these
pictures starting from the moment when \scri \ disappears,
and then to follow the evolution backwards in time. The result
is displayed in Figure 8; when read from right to left this
figure shows how the event horizon moves outward from the point at
the spatial origin where it is born.

\vspace{1cm}
\noindent
\parbox{\textwidth}{
\unitlength 1.00mm
\linethickness{0.4pt}
\begin{picture}(120.00,35.00)
\bezier{15}(10.00,25.00)(10.00,29.14)(12.93,32.07)
\bezier{15}(12.93,32.07)(15.86,35.00)(20.00,35.00)
\bezier{15}(30.00,25.00)(30.00,29.14)(27.07,32.07)
\bezier{15}(27.07,32.07)(24.14,35.00)(20.00,35.00)
\bezier{15}(10.00,25.00)(10.00,20.86)(12.93,17.93)
\bezier{15}(12.93,17.93)(15.86,15.00)(20.00,15.00)
\bezier{15}(30.00,25.00)(30.00,20.86)(27.07,17.93)
\bezier{15}(27.07,17.93)(24.14,15.00)(20.00,15.00)
\bezier{15}(40.00,25.00)(40.00,29.14)(42.93,32.07)
\bezier{15}(70.00,25.00)(70.00,29.14)(72.93,32.07)
\bezier{15}(100.00,25.00)(100.00,29.14)(102.93,32.07)
\bezier{15}(42.93,32.07)(45.86,35.00)(50.00,35.00)
\bezier{15}(72.93,32.07)(75.86,35.00)(80.00,35.00)
\bezier{15}(102.93,32.07)(105.86,35.00)(110.00,35.00)
\bezier{15}(60.00,25.00)(60.00,29.14)(57.07,32.07)
\bezier{15}(90.00,25.00)(90.00,29.14)(87.07,32.07)
\bezier{15}(120.00,25.00)(120.00,29.14)(117.07,32.07)
\bezier{15}(57.07,32.07)(54.14,35.00)(50.00,35.00)
\bezier{15}(87.07,32.07)(84.14,35.00)(80.00,35.00)
\bezier{15}(117.07,32.07)(114.14,35.00)(110.00,35.00)
\bezier{15}(40.00,25.00)(40.00,20.86)(42.93,17.93)
\bezier{15}(70.00,25.00)(70.00,20.86)(72.93,17.93)
\bezier{15}(100.00,25.00)(100.00,20.86)(102.93,17.93)
\bezier{15}(42.93,17.93)(45.86,15.00)(50.00,15.00)
\bezier{15}(72.93,17.93)(75.86,15.00)(80.00,15.00)
\bezier{15}(102.93,17.93)(105.86,15.00)(110.00,15.00)
\bezier{15}(60.00,25.00)(60.00,20.86)(57.07,17.93)
\bezier{15}(90.00,25.00)(90.00,20.86)(87.07,17.93)
\bezier{15}(120.00,25.00)(120.00,20.86)(117.07,17.93)
\bezier{15}(57.07,17.93)(54.14,15.00)(50.00,15.00)
\bezier{15}(87.07,17.93)(84.14,15.00)(80.00,15.00)
\bezier{15}(117.07,17.93)(114.14,15.00)(110.00,15.00)
\bezier{30}(46.38,34.13)(47.25,32.25)(50.00,31.96)
\bezier{30}(53.62,34.13)(52.75,32.25)(50.00,31.96)
\bezier{30}(40.87,28.62)(42.75,27.75)(43.04,25.00)
\bezier{30}(40.87,21.38)(42.75,22.25)(43.04,25.00)
\bezier{50}(12.78,17.93)(15.71,20.86)(15.71,25.00)
\bezier{50}(12.93,32.03)(15.86,29.10)(20.00,29.10)
\bezier{50}(27.07,32.03)(24.14,29.10)(20.00,29.10)
\bezier{50}(27.22,17.93)(24.29,20.86)(24.29,25.00)
\bezier{50}(27.22,32.07)(24.29,29.14)(24.29,25.00)
\bezier{50}(27.07,17.97)(24.14,20.90)(20.00,20.90)
\bezier{50}(12.93,17.97)(15.86,20.90)(20.00,20.90)
\bezier{50}(12.78,32.07)(15.71,29.14)(15.71,25.00)
\thicklines
\bezier{12}(42.90,26.16)(47.83,27.17)(48.84,32.10)
\bezier{12}(42.90,23.84)(47.83,22.83)(48.84,17.90)
\bezier{12}(57.10,26.16)(52.17,27.17)(51.16,32.10)
\bezier{12}(57.10,23.84)(52.17,22.83)(51.16,17.90)
\thinlines
\bezier{30}(46.38,15.87)(47.25,17.75)(50.00,18.04)
\bezier{30}(53.62,15.87)(52.75,17.75)(50.00,18.04)
\bezier{30}(59.13,28.62)(57.25,27.75)(56.96,25.00)
\bezier{30}(59.13,21.38)(57.25,22.25)(56.96,25.00)
\bezier{40}(74.44,33.33)(76.33,30.33)(80.00,30.33)
\bezier{40}(85.56,33.33)(83.67,30.33)(80.00,30.33)
\bezier{40}(88.33,19.44)(85.33,21.33)(85.33,25.00)
\bezier{40}(88.33,30.56)(85.33,28.67)(85.33,25.00)
\bezier{40}(71.67,30.56)(74.67,28.67)(74.67,25.00)
\bezier{40}(71.67,19.44)(74.67,21.33)(74.67,25.00)
\bezier{40}(85.56,16.67)(83.67,19.67)(80.00,19.67)
\bezier{40}(74.44,16.67)(76.33,19.67)(80.00,19.67)
\thicklines
\bezier{8}(80.00,30.33)(82.00,27.00)(85.33,25.00)
\bezier{8}(80.00,30.33)(78.00,27.00)(74.67,25.00)
\bezier{8}(80.00,19.67)(82.00,23.00)(85.33,25.00)
\bezier{8}(80.00,19.67)(78.00,23.00)(74.67,25.00)
\thinlines
\bezier{30}(106.22,25.00)(106.22,27.50)(105.22,29.78)
\bezier{10}(105.22,29.78)(104.56,31.20)(103.56,32.30)
\bezier{30}(106.22,25.00)(106.22,22.50)(105.22,20.22)
\bezier{10}(105.22,20.22)(104.56,18.80)(103.56,17.70)
\bezier{30}(110.00,21.22)(107.50,21.22)(105.22,20.22)
\bezier{10}(105.22,20.22)(103.80,19.56)(102.70,18.56)
\bezier{30}(110.00,21.22)(112.50,21.22)(114.78,20.22)
\bezier{10}(114.78,20.22)(116.20,19.56)(117.30,18.56)
\bezier{30}(113.78,25.00)(113.78,22.50)(114.78,20.22)
\bezier{30}(113.78,25.00)(113.78,27.50)(114.78,29.78)
\bezier{30}(110.00,28.78)(112.50,28.78)(114.78,29.78)
\bezier{30}(110.00,28.78)(107.50,28.78)(105.22,29.78)
\bezier{10}(114.78,20.22)(115.44,18.80)(116.44,17.70)
\bezier{10}(114.78,29.78)(115.44,31.20)(116.44,32.30)
\bezier{10}(114.78,29.78)(116.20,30.44)(117.30,31.44)
\bezier{10}(105.22,29.78)(103.80,30.44)(102.70,31.44)
\thicklines
\bezier{3}(110.00,26.50)(109.50,25.50)(108.50,25.00)
\bezier{3}(110.00,26.50)(110.50,25.50)(111.50,25.00)
\bezier{3}(110.00,23.50)(109.50,24.50)(108.50,25.00)
\bezier{3}(110.00,23.50)(110.50,24.50)(111.50,25.00)
\bezier{10}(50.00,10.00)(55.00,10.00)(60.00,10.00)
\thinlines
\put(61.00,10.00){\makebox(0,0)[lc]{= event horizon}}
\put(110.00,3.30){\makebox(0,0)[cc]{d}}
\put(20.00,3.00){\makebox(0,0)[cc]{a}}
\put(50.00,3.30){\makebox(0,0)[cc]{b}}
\put(80.00,3.00){\makebox(0,0)[cc]{c}}
\end{picture}

\smallskip
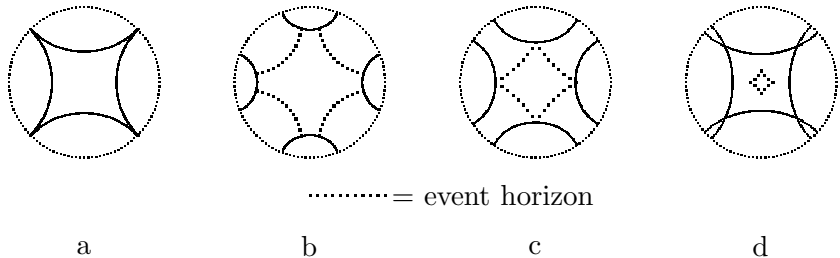
\noindent {\small Fig.~8: We follow the event horizon backwards
in time. The first slice (a) is at $t_P$. The next (b) is at $t = 0$,
the moment of time symmetry. Then the event horizon is a smooth
circle, and it remains smooth until the next earlier slice (c) when the
different pieces of the horizon meet each other at the
identification surfaces. Before that moment the horizon has
four kinks and is growing outward. The final slice (d) shows
the geometry just after
the event horizon is born as a point at the origin (at a time
$t_{\rm birth}$ when the asymptotic region does not exist).}
}

\vspace{2cm}
\noindent
\parbox{\textwidth}{
\unitlength 1.30mm
\linethickness{0.4pt}
\begin{picture}(97.07,27.46)(5,0)
\thicklines
\bezier{42}(70.27,20.00)(70.27,23.00)(74.29,25.22)
\bezier{46}(74.29,25.22)(78.17,27.46)(83.67,27.46)
\bezier{46}(70.27,20.00)(70.27,16.11)(74.29,13.55)
\bezier{52}(74.29,13.55)(77.64,11.36)(83.67,11.36)
\bezier{42}(97.07,20.00)(97.07,23.00)(93.05,25.22)
\bezier{46}(93.05,25.22)(89.17,27.46)(83.67,27.46)
\bezier{46}(97.07,20.00)(97.07,16.11)(93.05,13.55)
\bezier{52}(93.05,13.55)(89.70,11.36)(83.67,11.36)
\bezier{20}(83.67,18.00)(81.00,18.00)(78.87,18.93)
\bezier{20}(83.67,18.00)(86.34,18.00)(88.47,18.93)
\thinlines
\bezier{30}(78.87,18.93)(77.67,19.50)(77.00,20.50)
\bezier{30}(88.47,18.93)(89.67,19.50)(90.34,20.50)
\thicklines
\bezier{90}(77.54,20.00)(83.67,24.27)(89.80,20.00)
\bezier{12}(73.70,22.00)(73.70,24.17)(76.70,25.36)
\bezier{12}(76.70,25.36)(79.30,26.81)(83.70,26.81)
\bezier{12}(73.70,22.00)(73.70,19.49)(76.70,17.84)
\bezier{12}(76.70,17.84)(79.20,16.43)(84.20,16.82)
\bezier{10}(93.70,22.00)(93.70,24.17)(90.70,25.56)
\bezier{12}(90.70,25.56)(88.10,26.81)(83.70,26.81)
\bezier{12}(84.20,16.82)(87.67,17.67)(88.67,14.25)
\bezier{8}(88.67,14.25)(89.17,12.00)(87.57,11.60)
\thinlines
\bezier{10}(87.57,11.60)(85.67,12.00)(85.39,15.42)
\bezier{8}(85.39,15.42)(85.14,18.08)(82.89,18.00)
\thicklines
\bezier{9}(82.89,18.00)(80.39,17.70)(78.09,18.28)
\bezier{9}(78.09,18.28)(75.48,19.00)(74.97,21.50)
\bezier{6}(74.97,21.50)(74.73,23.17)(75.97,24.17)
\bezier{12}(75.97,24.17)(79.17,26.30)(83.64,26.30)
\bezier{12}(83.64,26.30)(87.14,26.30)(90.47,24.80)
\bezier{8}(90.47,24.80)(92.67,23.67)(92.39,21.92)
\bezier{10}(92.39,21.92)(91.56,20.00)(88.64,19.00)
\bezier{4}(92.88,20.17)(93.73,21.17)(93.70,22.00)
\bezier{8}(92.14,17.33)(91.17,18.42)(92.96,20.17)
\bezier{10}(92.23,13.00)(93.81,14.33)(92.14,17.33)
\thinlines
\bezier{12}(88.89,16.00)(91.06,12.42)(92.23,13.00)
\bezier{6}(88.64,19.00)(87.89,18.83)(88.89,16.00)
\thicklines
\bezier{42}(16.60,20.00)(16.60,23.00)(20.62,25.22)
\bezier{46}(20.62,25.22)(24.50,27.46)(30.00,27.46)
\bezier{46}(16.60,20.00)(16.60,16.11)(20.62,13.55)
\bezier{52}(20.62,13.55)(23.97,11.36)(30.00,11.36)
\bezier{42}(43.40,20.00)(43.40,23.00)(39.38,25.22)
\bezier{46}(39.38,25.22)(35.50,27.46)(30.00,27.46)
\bezier{46}(43.40,20.00)(43.40,16.11)(39.38,13.55)
\bezier{52}(39.38,13.55)(36.03,11.36)(30.00,11.36)
\bezier{20}(30.00,18.00)(27.33,18.00)(25.20,18.93)
\bezier{20}(30.00,18.00)(32.67,18.00)(34.80,18.93)
\thinlines
\bezier{30}(25.20,18.93)(24.00,19.50)(23.33,20.50)
\bezier{30}(34.80,18.93)(36.00,19.50)(36.67,20.50)
\thicklines
\bezier{90}(23.87,20.00)(30.00,24.27)(36.13,20.00)
\bezier{10}(32.67,16.78)(33.00,19.80)(35.22,17.70)
\bezier{12}(35.22,17.70)(36.67,15.89)(36.67,13.50)
\bezier{8}(36.67,13.50)(36.67,11.50)(34.44,12.00)
\bezier{8}(34.44,12.00)(32.67,13.56)(32.67,16.78)
\bezier{12}(20.03,22.00)(20.03,24.17)(23.03,25.36)
\bezier{12}(23.03,25.36)(25.63,26.81)(30.03,26.81)
\bezier{12}(20.03,22.00)(20.03,19.49)(23.03,17.84)
\bezier{12}(23.03,17.84)(25.53,16.43)(30.03,16.43)
\bezier{12}(40.03,22.00)(40.03,24.17)(37.03,25.36)
\bezier{12}(37.03,25.36)(34.43,26.81)(30.03,26.81)
\bezier{12}(40.03,22.00)(40.03,19.49)(37.03,17.84)
\bezier{12}(37.03,17.84)(34.53,16.43)(30.03,16.43)
\put(30.00,4.00){\makebox(0,0)[cc]{a}}
\put(84.00,4.00){\makebox(0,0)[cc]{b}}
\end{picture}

\smallskip
\noindent {\small Fig.~9:
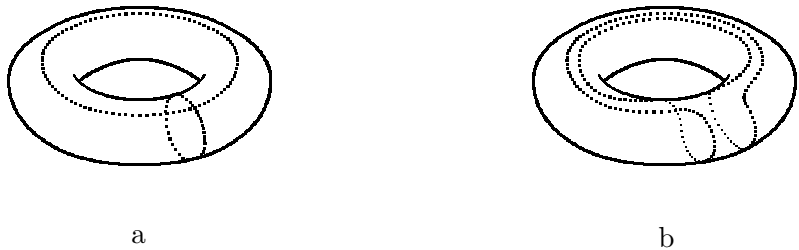
The event horizon (dotted curves) is born as two intersecting
circles on the torus (a). It is possible to find a spatial
slicing (b) such that the event horizon becomes a smooth curve a
moment after. However, the slice in which the horizon is
born cannot be smooth since the
caustics have a kink at the point where they intersect.
Hence this picture is topologically accurate but metrically
misleading.}
}
\vspace{5mm}

Note that the kinks in the event horizon are moving outward
faster than light; otherwise expressed, the event horizon grows
from a pair of spacelike caustics. If we consider slices of
equal stereographic time ${\tau}$ instead we find that the
event horizon has a different spatial topology in its early
stages. At first sight
this is a bit confusing (it is also reminiscent of the changing
spatial topology of the event horizon found recently in
numerical simulations \cite{Teukolsky}). To see what goes on
it is helpful to remember that space is really a torus with
one asymptotic region. Then we can draw a faithful picture
of the caustics at the bottom of the event horizon, and
observe how the event horizon becomes smooth a moment
after. See Figure 9, which then corresponds to a very
particular slicing of our spacetime. After some reflection,
one realizes that by changing the slicing a little bit
a variety of spatial topologies for the event horizon can
be obtained. This is however a slicing dependent statement
of no particular importance.

Let us sketch the complications which occur in the general
case. First the toroidal wormhole can be generalized by choosing a
less symmetric fundamental region. This case is somewhat more
difficult to analyze---a little care is needed to locate the
event horizon. Nevertheless there are only very modest
changes in the conclusions, and for this reason we do not give
the details here. (The only noteworthy change is that the
caustics no longer have to intersect in a single point.)
For higher genus wormholes the situation is again very
similar to the simple case that we have studied. The
main conclusions are that it remains true in all cases that
the domain of outer communication is isometric to the BTZ
exterior (perhaps this is clearest in the ``trousers''
construction of the initial data), and that the
black holes will be non-eternal since the event horizon is
born at a time later than the birth of the universe in
all cases.

To sum up the discussion so far, our black holes are not eternal.
At early times the length of the event horizon is growing
due to the presence of caustics. Indeed although the domain
of outer communication is always isometric to the BTZ exterior
(and hence static), the spacetime as a whole admits no global
Killing vector at all. Nor does its exterior, which is larger
than its domain of outer communication.

\vspace{5ex}
\noindent
{\large VI. Topological Censorship.}

\vspace{3ex}

\noindent Our black hole can be described as the
creation and collapse
of a wormhole, or as a topological geon. A topological geon
is a spacetime with a localized configuration having a
non-Euclidean topology. Whenever there is a wormhole, a
question concerning causality suggests itself: Is it perhaps
possible to travel almost instantaneously through the wormhole
between two points that are otherwise widely separated from
each other? In four spacetime dimensions there are precise
theorems that are relevant to this question. In particular
it is known that an asymptotically flat spacetime with a
non-simply connected Cauchy surface necessarily has a
singular time evolution \cite{Gannon}, so that the time
available to send signals through a wormhole is limited. A
very precise statement about this is the topological
censorship theorem \cite{Friedman}, which states that all
causal curves connecting past and future null infinity are
homotopic to a curve that lies in a simply connected
neighborhood of null infinity. In this sense it is
impossible to probe the interior topology actively from
far away. Sufficient conditions for the theorem to hold are
asymptotic flatness, global hyperbolicity and the averaged
null energy condition. A close relative of the theorem
states that the domain of outer communication is simply
connected. Note that the theorem does not
entail the stronger notion of passive topological
censorship, according to which the topology is unobservable
from infinity. Indeed there are counterexamples to passive
topological censorship.

Now the question is to which extent our wormhole agrees with
these notions. Since our spacetime is asymptotically anti-de Sitter
and not globally hyperbolic the answer is not quite
obvious. Moreover \scri \ has the topology of a cylinder and
is not simply connected. Hence the domain of outer communication
is not simply connected, not even for the BTZ black hole.
On the other hand we know that the domain of outer communication
is isometric to that of the BTZ black hole for all our
examples, so that no further topological complications can
be actively probed from infinity. In this sense, active
topological censorship does hold for all our wormholes.
A direct way to see what
happens if one tries to send a signal from \scri \
around the wormhole and back to \scri \
is to choose a path that starts in one of the openings of the
``tent'' and ends at the opposite side. The minimum amount of
sausage time required for such a trip is ${\pi}$, but we have
seen that \scri \ never lasts that long.

\noindent
\parbox{\textwidth}{
\unitlength 1.00mm
\linethickness{0.4pt}
\begin{picture}(108.72,59.00)(10,15)
\put(56.00,54.00){\line(1,-1){23.00}}
\put(102.00,54.00){\line(-1,-1){23.00}}
\bezier{170}(56.00,54.00)(64.06,40.06)(56.00,26.00)
\bezier{170}(102.00,54.00)(93.94,40.06)(102.00,26.00)
\thicklines
\bezier{116}(56.00,54.00)(67.00,48.00)(79.00,59.00)
\bezier{116}(102.00,54.00)(91.00,48.00)(79.00,59.00)
\bezier{116}(56.00,26.00)(67.00,32.00)(79.00,21.00)
\bezier{116}(102.00,26.00)(91.00,32.00)(79.00,21.00)
\put(79.00,48.00){\makebox(0,0)[cc]{Black Hole}}
\put(79.00,44.00){\makebox(0,0)[cc]{Region}}
\bezier{30}(59.00,48.00)(69.00,38.00)(79.00,28.00)
\bezier{30}(79.00,28.00)(89.00,38.00)(99.00,48.00)
\thinlines
\put(62.00,51.50){\makebox(0,0)[cc]{*}}
\put(69.00,52.22){\makebox(0,0)[cc]{*}}
\put(75.00,55.22){\makebox(0,0)[cc]{*}}
\put(96.00,51.50){\makebox(0,0)[cc]{*}}
\put(89.00,52.22){\makebox(0,0)[cc]{*}}
\put(83.00,55.22){\makebox(0,0)[cc]{*}}
\put(62.00,27.47){\makebox(0,0)[cc]{*}}
\put(69.00,26.75){\makebox(0,0)[cc]{*}}
\put(75.00,23.75){\makebox(0,0)[cc]{*}}
\put(96.00,27.47){\makebox(0,0)[cc]{*}}
\put(89.00,26.75){\makebox(0,0)[cc]{*}}
\put(83.00,23.75){\makebox(0,0)[cc]{*}}
\put(49.49,42.31){\vector(4,-1){9.49}}
\put(49.23,42.31){\makebox(0,0)[rc]{\scri}}
\put(108.46,42.31){\vector(-4,-1){9.49}}
\put(108.72,42.31){\makebox(0,0)[lc]{\scri}}
\end{picture}

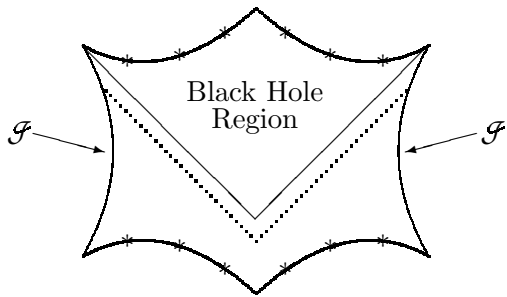
\noindent {\small Fig.~10: A vertical slice through the
stereographic image of the ``tent'', centered on its center of
symmetry. The solid light rays bordering the black hole region
are the event horizon. The dotted lines are two light rays that reach
\scri \ in such a way that they violate passive topological
censorship.}
}

\vspace{5mm}

Nevertheless our black holes can be distinguished from the
BTZ black hole, also from far away. To see how, choose a
point on the ``central line'' going vertically through
the fundamental region (i.e. on the line ${\rho} = 0$
if sausage coordinates are used) and
to the past of the event horizon. Then connect it
to \scri \ by means of two light rays that end in different openings
of the ``tent'', which means that the light rays
form a path that winds around the wormhole. (This
is illustrated in Figure 10.) In other
words we can see the full extent of the wormhole from infinity,
even though we cannot travel through it. Therefore passive
topological censorship does not hold for our wormholes.

\vspace{5ex}
\noindent
{\large VII. Apparent Horizons and Trapped Surfaces.}

\vspace{3ex}

\noindent In this section we venture into the interior of our black
holes in order to look for trapped curves. To define this notion,
consider the two families of future directed null geodesics that are
orthogonal to a closed curve in space. (If we were in 3+1 dimensions,
we would have to define trapped surfaces instead, with appropriate
changes in the definitions that follow.) If both families of
light rays converge, there is clearly something non-trivial going
on, and the curve is said to be trapped. It is an essential part
of the definition that the curve be closed.
The definition is important because there are singularity theorems that say
that a singularity necessarily occurs to the future of an initial data
surface that contains a trapped curve (although it may happen that
the singularity is avoided if closed timelike curves
appear \cite{Claire}---a remark that is relevant to the spinning
BTZ black hole). The theorems can also be proved under the
assumption that an outer trapped curve occurs. An outer trapped
curve (in a given spatial slice) is defined as the boundary of
a two dimensional manifold-with-boundary whose outgoing
light rays converge \cite{Wald}. Note that a trapped curve need
not be outer trapped since a trapped curve need not be the
boundary of any region---trapped curves are allowed to
intersect themselves, while outer
trapped ones cannot do so. If the (outgoing) light rays have zero
convergence the curve is said to be marginally (outer) trapped.
Given a foliation of spacetime, a spatial region sitting
inside an outer marginally trapped curve is called a trapped region,
and the apparent horizon is ordinarily defined as the boundary of
the union of all trapped regions. The conventional definition of
the apparent horizon therefore demands that spacetime has been
sliced into spatial hypersurfaces, it also depends strongly on this particular
slicing (see eg reference \cite{slicing} for a striking example).
However, since our spacetimes are simple enough that we can find all the
marginally
trapped curves, we can afford to define
the apparent horizon as the boundary of the spacetime region in
which the trapped curves appear. Hence to us both the apparent
horizon and the event horizon are spacetime concepts.

Let us consider the BTZ black hole first, because it is a
particularly transparent case. To construct trapped curves in this
case we just place vertices of two backward light cones anywhere
on the line  that represents the singularity. See Figure 11, where we employ
stereographic coordinates so that lightcones appear as
ordinary cones. The intersection of these light
cones forms a flow line of the Killing vector $J_{TY}$ that
we use to identify points in anti-de Sitter space, and therefore
it will be turned into a smooth closed curve by the identification.
By construction both families of orthogonal light rays converge
(namely to the vertices of the two light cones), and therefore
this curve is trapped. Clearly there will be such a trapped
curve through any point in the interior of the BTZ black hole.
The event horizon is marginally trapped, since in that case the
outgoing light rays lie on a null plane (with its vertex on the
boundary of adS) and have zero convergence. In any spatial
slicing the event horizon forms the boundary of the trapped
region, and therefore the apparent horizon coincides with the
event horizon.

\vspace{1cm}
\noindent
\parbox{\textwidth}{
\unitlength 1.00mm
\linethickness{0.4pt}
\begin{picture}(86.00,43.42)(-5,2)
\bezier{8}(50.00,40.00)(50.00,41.42)(52.93,42.42)
\bezier{8}(52.93,42.42)(55.86,43.42)(60.00,43.42)
\bezier{8}(70.00,40.00)(70.00,41.42)(67.07,42.42)
\bezier{8}(67.07,42.42)(64.14,43.42)(60.00,43.42)
\bezier{8}(50.00,40.00)(50.00,38.58)(52.93,37.58)
\bezier{8}(52.93,37.58)(55.86,36.58)(60.00,36.58)
\bezier{8}(70.00,40.00)(70.00,38.58)(67.07,37.58)
\bezier{8}(67.07,37.58)(64.14,36.58)(60.00,36.58)
\put(51.00,38.50){\line(6,1){18.00}}
\put(54.00,39.00){\line(-1,-1){19.00}}
\put(85.00,22.00){\line(-1,1){19.00}}
\bezier{20}(37.50,20.00)(50.00,33.00)(50.00,40.00)
\bezier{28}(34.20,18.00)(34.20,20.80)(40.00,22.79)
\bezier{20}(40.00,22.79)(45.80,24.77)(54.00,24.77)
\bezier{16}(73.80,18.00)(73.80,20.80)(68.00,22.79)
\bezier{20}(68.00,22.79)(62.20,24.77)(54.00,24.77)
\bezier{44}(34.20,18.00)(34.20,15.20)(40.00,13.21)
\bezier{66}(40.00,13.21)(45.80,11.23)(54.00,11.23)
\bezier{20}(73.80,18.00)(73.80,15.20)(68.00,13.21)
\bezier{66}(68.00,13.21)(62.20,11.23)(54.00,11.23)
\bezier{11}(46.20,20.00)(46.20,22.80)(52.00,24.79)
\bezier{11}(52.00,24.79)(57.80,26.77)(66.00,26.77)
\bezier{22}(85.80,20.00)(85.80,22.80)(80.00,24.79)
\bezier{11}(80.00,24.79)(74.20,26.77)(66.00,26.77)
\bezier{16}(46.20,20.00)(46.20,17.20)(52.00,15.21)
\bezier{20}(52.00,15.21)(57.80,13.23)(66.00,13.23)
\bezier{44}(85.80,20.00)(85.80,17.20)(80.00,15.21)
\bezier{60}(80.00,15.21)(74.20,13.23)(66.00,13.23)
\thicklines
\bezier{88}(67.83,13.50)(63.17,31.17)(60.00,32.67)
\bezier{60}(60.00,32.67)(58.00,34.83)(52.17,24.83)
\thinlines
\put(66.00,41.00){\line(-1,-1){7.00}}
\bezier{32}(59.00,34.00)(53.50,28.50)(48.00,23.00)
\bezier{20}(82.50,20.00)(70.00,33.00)(70.00,40.00)
\put(59.83,32.83){\vector(3,4){3.75}}
\put(62.83,28.67){\vector(1,4){2.25}}
\put(54.00,39.00){\line(1,-1){7.00}}
\bezier{34}(60.00,33.00)(66.00,27.00)(72.00,21.00)
\end{picture}

\noindent {\small Fig.~11:
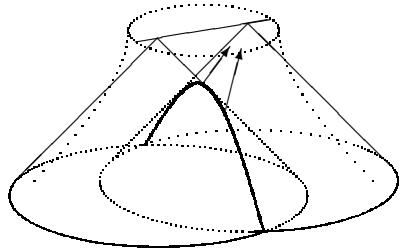
Trapped curves in the BTZ black holes can be
obtained as the intersection (heavily drawn hyperbola)
of pairs of light cones whose
vertices have been placed on the singularity. Since we use
stereographic coordinates these light cones appear as ordinary
cones in the picture.  The outer, dotted hyperboloid represents infinity.
The two converging arrows are lightlike normals
on one side of the intersection; a similar pair exists on the other side,
along the left light cone. This shows that the hyperbola is trapped.}
}
\vspace{5mm}

Let us now consider the wormhole, and more particularly the
simplest wormhole as defined in section V. The first
observation is that any curve that lies in the event horizon
and avoids the caustics is by construction marginally outer
trapped, since the family of orthogonal outgoing light rays
lies on the backwards light cone of the last point on \scri .
By deforming such a curve inward it is possible to find
outer trapped curves in the interior. Therefore the main
conclusion is again that the interior of the black hole is
trapped, and that the apparent horizon coincides with the
event horizon. (In 3+1 dimensions it is possible
to construct a black hole which is locally isometric with
anti-de Sitter space but for which this property does not
hold \cite{HP}.) A minor remark is that it is essential to
require that the outer trapped curve bound a region:
it is possible to find closed self intersecting curves for
which one family of outgoing light rays converge and which
extends beyond the event horizon. If such curves were
admitted the theorem that the apparent horizon cannot
lie outside the event horizon would not be true.

It is more interesting to look for curves that are trapped
in both directions. Consider first the time symmetric initial
data surface $U = 0$, and think of its covering space as a Poincar\'{e}
disk.
A geodesic will be closed provided that it
meets the boundary of the disk at the fixed points of some
transvection belonging to the isometry group
which is used to identify points. Although the discrete
isometry group has only two generators it has an infinite
number of elements; indeed to every ``word'' in the
generators $a$ and $b$ and their inverses
there corresponds an element of
${\Gamma}$. Hence there are very many such
fixed points, and indeed it is not difficult to convince
oneself  (by means of a tesselation of the disk into
copies of the fundamental region)
that the closed geodesics fill the interior of the
black hole densely, while there are none in the exterior.

We can also think of the initial data surface as a
hyperboloid embedded in Minkowski space (compare Fig.~1a---stereographic
projection is essentially an isometry for this surface). Then
the geodesics are given by the intersection of the hyperboloid
with some timelike plane through the origin; analytically

\begin{equation} x_0X + y_0Y - {\tau}_0T = 0 \hspace{5mm} ;
\hspace{1cm} x_0^2 + y_0^2 - {\tau}_0^2 = 1 \ . \end{equation}

\noindent  But this is also the
intersection of the initial data surface with two lightlike
planes in adS, viz.

\begin{equation} x_0X + y_0Y - {\tau}_0T \mp U = 0 \ .
\end{equation}

\noindent Such a plane is the backwards light cone of
a point on the boundary of adS having stereographic
coordinates $\pm(x_0, y_0, {\tau}_0)$. Any such point
on the boundary will have an image on the boundary of
the fundamental region, at which the corresponding light
rays in the black hole spacetime end. We can therefore
conclude that any closed geodesic on the initial data
surface is a marginally trapped curve in the black hole
spacetime; the interior of the black hole in the initial
data slice is densely covered with marginally trapped curves.
Most but by no means all of them are self intersecting.

To the future of the marginally trapped curves genuinely
trapped curves will appear. To see how this happens,
consider an arbitrary closed geodesic in the initial data
surface. By means of a global isometry of covering space it
can be brought to coincide with a straight line through the
center of the disk. It is a flow line of a unique transvection
belonging to ${\Gamma}$ and having a line of fixed points
in anti-de Sitter space. This line coincides with
the singularity of some BTZ spacetime (namely, the
one whose group is generated by that unique
transvection). Hence we can repeat the construction
that gave us trapped curves
there. Finally we undo the global isometry, and conclude that
the trapped curve arises as the intersection of a pair
of light cones suspended from a line of fixed points which
(in stereographic coordinates) is precisely the straight line
joining the points $- (x_0, y_0, {\tau}_0)$ and $(x_0, y_0,
{\tau}_0)$.

We have now constructed all marginally trapped curves. To see
this it is enough to see that all families of light rays
that have zero convergence have to converge to a point on
\scri . This is indeed so in 2 + 1 dimensions where there
can be no shear to complicate matters. To ensure that a
spatial curve through such a null
plane is closed it is necessary that its vertex lie on a
fixed point of an element of the discrete isometry
group that is used to identify points.
But we have already accounted for the
intersections of all such null planes. The generalization
to more complicated wormhole spacetimes is straightforward
and is omitted here.

\newpage
\noindent
{\large VIII. Conclusions and Open Questions.}

\vspace{3ex}

\noindent In this paper we have constructed all black holes which
can be obtained by taking the quotient of a region of 2+1
dimensional anti-de Sitter space, which have one asymptotic
region only, and whose fundamental group can be fully characterized
by time symmetric initial data. Their event horizons and their
trapped regions were displayed and it was shown in what sense topological
censorship holds.
 The event horizons are not eternal,
and have caustics which were described in some detail. Our account
of the marginally trapped surfaces was complete. In general trapped
surfaces are not so easy to find, so this clearly shows the
advantages of working in 2+1 dimensions.

The restriction to one asymptotic region can easily be lifted (see
ref. \cite{Brill1} for some possibilities in this direction). The
restriction to time symmetric initial data on the other hand is less
easily dispensed with. We accept a solution only if
it has no naked singularities. For the time symmetric initial data we
were able to analyze this requirement in detail, but we have been
unable to make any progress in the general case. It is even conceivable
that there are no further acceptable solutions. On the other hand, it
may seem reasonable to expect that there should be a ``spinning'' wormhole
as well. The BTZ black hole indeed admits a generalization with spin
\cite{BHTZ}. However, the nature of the singularities of this spacetime
is quite different from that of the spinless BTZ black hole---in the
spinning case there are no fixed point singularities, only (hidden)
regions with closed timelike curves. As shown in ref.~\cite{HP} a
complete analysis of the causal structure in such situations is not
straightforward. In particular the fundamental region---which
was so useful in the case studied in the present paper---is no longer
such a useful tool in the more general case.
The spinning case is therefore open, and we have no final answer to
the question of how many black holes with one asymptotic region there
may be.

Another open question concerns quantum theory. We have nothing to say
about this here, but we wish to stress that the motivation for studying
black holes in 2+1 dimensions is not primarily to find instructive
examples of the behavior of event horizons and the like---even though
we have probably made it clear that we do find this interesting in
itself. The primary motivation is that one may hope that 2+1 dimensional
gravity can play the same role for quantum gravity that the Schwinger
model played for QCD. We believe that the solutions described in
this paper deserve study from this point of view.

\vspace{5ex}
\noindent
\underline{Acknowledgement:}

\vspace{3ex}

\noindent We thank Milagros Izquierdo for lessons in mathematics. IB and PP
were supported by the NFR.

\newpage
\noindent
{\large  Appendix A}

\vspace{3ex}

\def\theequation{A.\arabic{equation}}
\setcounter{equation}{0}
\noindent In the stereographic description---which was described more
fully in ref. \cite{Brill2}---we
project from one point in the embedded adS space to the tangent
space T of the antipodal point. This tangent space inherits a
flat, Lorentzian metric from the 4-dimensional embedding space. If we
choose the center of projection at $(T, U, X, Y) = (0,- 1, 0, 0)$
and the tangent space T at $U = 1$ with coordinates
$x, y, \tau = X,Y,T$, this metric is simply

\begin{equation} ds_{\rm flat}^2 = -d\tau^2 + dx^2 + dy^2 \ .
\end{equation}

\noindent The projection of a point $X^\mu$ on the adS space (3)
to a point $x^\mu$ in T is described by

\begin{equation}  x^\mu = {2 X^\mu\over U + 1}
\qquad (X^\mu \neq U) \ , \end{equation}

\noindent which defines another metric on T
(corresponding to the adS geometry),

\begin{equation} ds_{\rm adS}^2 = \left({1\over 1-r^2}
\right)^2(-d\tau^2+dx^2+dy^2) \qquad
{\rm where} \qquad r^2={-\tau^2+x^2+y^2\over 4} \ . \end{equation}

\noindent Clearly the relation between the two metrics is
{\it conformal}---as
in the Euclidean context, stereographic projection is a conformal map.

This map does not cover all of adS space; we say that it is ``centered on''
the point antipodal to the center of projection (for the projection of
Eq (16) the center is the point $(0,1,0,0)$).
Since a projection can be
centered on any point, all of adS space can be covered
by an atlas of such maps. The part of adS space which is
covered by the stereographic
projection is given by $U > - 1$ and this region
is mapped into the interior of a hyperboloid, $x^2 + y^2 - \tau^2 < 4$.

The isometries of adS space become certain conformal
isometries of T that are in general not Lorentz transformations of
the Minkowski metric; but those adS transformations that leave the origin
of T fixed (and hence leave $r$ constant) {\it are}
also Lorentz transformations of the flat (inherited) metric of T.
Thus any isometry of adS that has fixed points can be represented as a Lorentz
transformation in a suitably centered stereographic projection. The
projection centered on $(0,1,0,0)$ is particularly suitable for
our purposes since the Killing vectors that take the surface $U = 0$
into itself assume their familiar Minkowski
space form (as generators of the three dimensional Lorentz
group) in terms of the coordinates $x$, $y$ and ${\tau}$.

\newpage
\noindent
{\large Appendix B}
\def\theequation{B.\arabic{equation}}
\setcounter{equation}{0}

\vspace{3ex}

\noindent The sausage coordinates were fully described in earlier papers
\cite{Holst}, so we can be brief. The idea is to introduce a time
coordinate $t$ such that

\begin{equation} T = V\cos{t} \hspace{1cm} U = V\sin{t} \ .
\end{equation}

\noindent Then the equation for the anti-de Sitter hyperboloid and
its metric become respectively

\begin{equation} - V^2 + X^2 + Y^2 = - 1 \end{equation}

\begin{equation} ds^2 = dX^2 + dY^2 - dV^2 - V^2dt^2 \ . \end{equation}

\noindent For constant $t$ these are the equations that define the
hyperbolic plane (embedded in Minkowski space). We arrive at the
sausage coordinates $t, {\rho}, {\phi}$ if we introduce Poincar\'{e}'s
stereographic coordinates on the hyperbolic plane itself. The sausage
coordinates are related to the embedding coordinates by

\begin{equation} X = \frac{2{\rho}}{1 - {\rho}^2}\cos{\phi} \hspace{1cm}
Y = \frac{2{\rho}}{1 - {\rho}^2}\sin{\phi} \end{equation}

\begin{equation} T = \frac{1 + {\rho}^2}{1 - {\rho}^2}\cos{t}
\hspace{1cm} U = \frac{1 + {\rho}^2}{1 - {\rho}^2}\sin{t} \ .
\end{equation}

\noindent The metric in these coordinates is

\begin{equation} ds^2 = - \left( \frac{1 + {\rho}^2}{1 - {\rho}^2}
\right)^2dt^2 + \frac{4}{(1 - {\rho}^2)^2}(d{\rho}^2 + {\rho}^2
d{\phi}^2) \ . \end{equation}

A useful fact is that it takes an amount of sausage time $t = {\pi}/2$
for a light ray to go from the origin to \scri . A word of warning concerns
pictures such as our Figure 8. The point is that the Killing vectors
that we use to generate the identifying isometries do not lie in the
disks of constant $t$, except for the particular disk at $t = 0$. In
Figure 8, disks at different $t$ values are shown, and it may appear as
if the event horizon has kinks at the boundaries of the fundamental
region. This is not so; the event horizon is indeed smooth, but the
disks are not (except when $t = 0$).

\end{document}